\documentclass{article}
\usepackage{arxiv}

\usepackage{graphicx}
\usepackage[utf8]{inputenc} % allow utf-8 input
\usepackage[T1]{fontenc}    % use 8-bit T1 fonts
\usepackage{hyperref}       % hyperlinks
\usepackage{url}            % simple URL typesetting
\usepackage{booktabs}       % professional-quality tables
\usepackage{amsfonts}       % blackboard math symbols

\usepackage{nicefrac}       % compact symbols for 1/2, etc.
\usepackage{microtype}      % microtypography
\usepackage{lipsum}

\usepackage{bm}
\usepackage{amsmath} %must be before cases or error
\usepackage{cases}

\usepackage{mathtools}

\title{Full-field stress computation from measured deformation fields: a hyperbolic formulation}

\author{
 Benjamin C. Cameron\\
  Department of Civil and Environmental Engineering\\
  Massachusetts Institute of Technology\\
  \texttt{bcam@mit.edu} \\
  \AND
   C. Cem. Tasan\\
   Department of Materials Science and Engineering\\
   Massachusetts Institute of Technology\\
   \texttt{tasan@mit.edu} \\
}

\begin{document}

\maketitle
\begin{abstract}
Recent developments in microscopic imaging techniques and correlation algorithms enable measurement of strain fields on a deforming material at high spatial and temporal resolution. In such cases, the computation of the stress field from the known deformation field becomes an interesting possibility. This is known as an inverse problem. Current approaches to this problem, such as the finite element update method, are generally over-determined and must rely on statistical approaches to minimize error. This provides approximate solutions in some cases, however, implementation difficulties, computational requirements, and accuracy are still significant challenges. Here, we show how the inverse problem can be formulated deterministically and solved exactly in two or three dimensions for large classes of materials including isotropic elastic solids, Newtonian fluids, non-Newtonian fluids, granular materials, plastic solids subject to co-directionality, and some other plastic solids subject to associative or non-associative flow rules. This solution is based on a single assumption of the alignment of the principal directions of stress and strain or strain rate. No further assumptions regarding incompressibility, pressure independence, yield surface shape or the hardening law are necessary. This assumption leads to a closed, first order, linear system of hyperbolic partial differential equations with variable coefficients. The solution of this class of problems is well established and hence the equations can be solved to give the solution for any geometry and loading condition, enabling broad applicability to a variety of problems. We provide a numerical proof-of-principle study of the plastic deformation of a two-dimensional bar with spatially varying yield stress and strain hardening coefficient. The results are validated against the solution of the corresponding forward problem - solved with a commercial finite element solver - indicating the solution is exact up to numerical error (the normalized root mean square error of the stress is $1.63\times10^{-4}$).  No model calibration or material parameters are required. The sensitivity of the solution to error in the input data is also analyzed. Interestingly, this solution procedure lends itself to a simple physical interpretation of stress propagation through the material. Finally, we provide some examples showing how this approach may be analytically applied to both solid and fluid mechanics problems.
\end{abstract}

% keywords can be removed
\section{Introduction}
\label{sec:intro}
\begin{figure}[ht!]
\centering
\includegraphics[width=1\textwidth]{{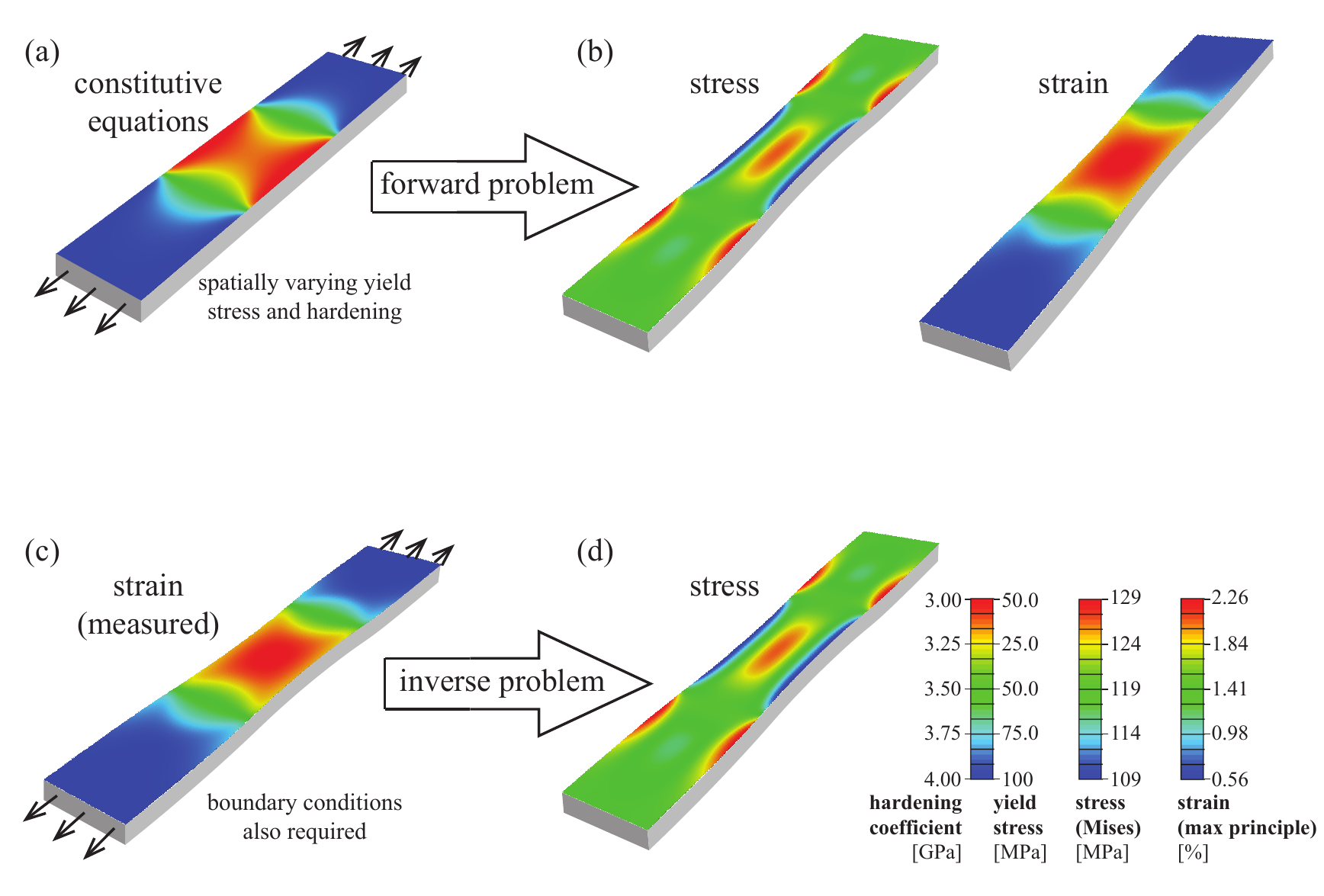}}
\caption{Comparison of the forward problem and inverse problem for the case of a plastically deforming bar with spatially varying yield stress and hardening coefficient. For the forward problem, (a) the constitutive equations and boundary conditions are used to compute (b) the stress and strain. For the inverse problem, (c) the measured strain and boundary conditions are used to compute (d) the stress.}\label{fig:inverse}
\end{figure}

Numerous experimental approaches can be used to measure deformation fields in two and three dimensions such as digital image correlation (DIC) \cite{Pan2009,Kang2005,Yan2015}, digital volume correlation \cite{Lenoir2007,Roux2008,Tudisco2015}, and particle image velocimetry \cite{Adrian2005,Heays2014,Elsinga2006}. However, it is often critical to know the stress field in addition to the deformation as it is the stress-deformation relationship that ultimately governs material behavior. Methods to directly measure the stress field can be more experimentally challenging. Approaches such as photo-elasticity can only be applied to the elastic deformation of dielectric materials with specific optical properties \cite{Ramesh2000}, residual stress measurements are often destructive \cite{Korsunsky2009}, and in-situ stress transducers have low spatial resolution \cite{Harris1994}. This experimental difficulty motivates the inverse problem: solving for the stress when the deformation is known and the constitutive relations of the material are unknown. The forward problem, where the constitutive equations are known but the deformation field is unknown, has been extensively researched since the pioneering work by Hill \cite{Hill1950} (Fig. \ref{fig:inverse}), leading to the development of advanced theories such as crystal plasticity \cite{Roters2010,Asaro1983} and strain gradient plasticity \cite{Fleck1994, Gurtin2005}. This approach, however, cannot be easily inverted to deal with scenarios where the constitutive equations remain unknown.

Note that the inverse problem has exact solutions in some cases with high degrees of symmetry. One example is static tension applied along the length of a bar with a constant cross section. Here, the stress is constant along the bar regardless of any complexity or variability in the constitutive equations. Other examples include the triaxial test \cite{Colliat-Dangus1988} and the bulge test \cite{Plancher2019}. It is these solutions that enable the measurement of constitutive equations for a range of materials because here both the stress and deformation can be known. However, this approach is limited due to several reasons: first, as soon as the symmetries break down the local stresses become unknown (e.g. when the necking occurs in a bar), second, one experiment generally only corresponds to one stress or deformation path, and the full constitutive equations require multiple deformation paths \cite{Plancher2019}, third, it does not apply when the material has heterogeneous properties, fourth, it does not apply when the geometry or loading condition gives rise to a heterogeneous stress field.

To solve the problem more generally, the primary approach taken in literature is to search for a stress field consistent with force equilibrium, the observed strain field, and constitutive equation form (e.g. linear isotropic elasticity). However, these problems are generally over determined and there is no solution which satisfies all requirements. For example, if the observed strain field deviates slightly from a solution determined by linear elasticity, and if linear elasticity is assumed, the exact solution will be impossible. To resolve this, current approaches drop or weaken one of these requirements so there are multiple possible solutions. This requirement is then replaced with some error metric that is minimized.

In the finite element update method, for example, the form of the constitutive equations is assumed, force equilibrium is solved, and the material parameters are updated to minimize the difference between the computed and observed strain \cite{Rethore2010,Siddiqui2017,Viala2018}. The advantage of this method is that it can be used for complex constitutive equations, such as crystal plasticity. An alternative approach is the equilibrium gap method. Here, instead of the force equilibrium being exact and minimizing the error in the strain, the strain is exact and the error is minimized in the force equilibrium (equilibrium gap) \cite{Crouzeix2009,Florentin2010}. Similarly, the virtual field method also minimizes the error in the force equilibrium \cite{Grediac2002,Pierron2012}. The constitutive equation gap method minimizes the error between the stress computed with assumed constitutive equations and a stress consistent with force balance \cite{Florentin2011}.

These approaches, and similar approaches (e.g. \cite{Gelin1994}), have enabled some quantitative estimation of material parameters, however the performance of these algorithms has been inconsistent (with errors typically ranging between 2 and 70 percent). For example, Florentin and Lubineau use the constitutive equation gap method to locally estimate the Young’s modulus of a material with a maximum error of approximately 10 percent \cite{Florentin2010}. However, there can also be significantly higher errors depending on the problem. For example, for a different problem analyzed by Florentin and Lubineau, the maximum error in the Young’s modulus was 43 percent. This was increased to 68 percent when the equilibrium gap method was used \cite{Florentin2010}. An additional challenge is that these generally become more computationally intensive and error prone with increasing complexity of the problem \cite{Grediac2004}. For example, a large number of parameters must be optimized over when the constitutive equations vary. Furthermore, there can be uncertainties regarding the uniqueness of the solution, and it is often unclear what form of constitutive equation must be assumed. Nevertheless, these approaches can produce quantitative results for some two dimensional problems with isotropic elasticity, and have even been extended to cases of elasto-plasticity with uniform material parameters \cite{Grediac2006}. For a more extensive review of these methods, see \cite{Bonnet2005, Hild2006,Avril2008} and references therein.

This article shows that by making more limited assumptions about constitutive equation, one can arrive at a closed, mathematically determinate, set of equations consistent with force balance and the observed strain field. The key assumption made is the alignment of the principal directions of stress with either the strain or the strain rate. We refer to this as the alignment assumption. Our approach makes specifying the form of the constitutive equation unnecessary and one can avoid multiple commonly made assumptions such as incompressibility, pressure independent behavior, yield stress value, hardening law, or yield surface shape. However, unlike approaches such as the finite element update method, this cannot be applied to in cases which violate the alignment assumption, such as crystal plasticity, without significant modifications. Combining this assumption of alignment of principal directions with force balance gives rise to a closed, three dimensional, first order, linear system of hyperbolic partial differential equations with variable coefficients (\S \ref{sec:theory}), the solution of which is well established and enables the stress to be directly computed for any geometry and loading condition.

This article is structured as follows. First, the governing equations are derived for small strain and finite strain visco-plasticity and elasticity (\S \ref{sec:theory}). We provide a numerical proof-of-principle study showing our approach gives accurate results for a bar with yield stress and hardening coefficient spatially varying in two dimensions (\S \ref{sec:proof_of_principle}). This solution procedure has a simple physical interpretation of stress propagating through the material which gives insight into how the boundary conditions should be specified (\S \ref{sec:stress_propagation}). We discuss more extensively for what material classes our approach can and cannot be applied (\S \ref{sec:range_of_applicability}). Finally, this solution procedure may also be applied analytically, hence, we give examples for both solids and fluids (\S \ref{sec:examples}). 

\section{Governing equations}
\label{sec:theory}
\subsection{Alignment assumption}
The key assumption made is the alignment assumption: that the principal directions of the Cauchy stress $\mathbf{T}$ are the same as either the strain or strain rate. This assumption was first made by saint-Venant in 1871 \cite{Timoshenko1983} and, though it is generally not made explicitly, it is implicit in multiple commonly used assumptions for elastic, plastic and viscous deformation. For example, consider the co-directionality assumption which specifies that the principle directions of the stress and strain rate are the same but also imposes requirements on ratios of the principal values \cite{VonMises1913,Levy1891,Prandtl1924,Reuss1930}. This co-directionality assumption ensures the equations are thermodynamically consistent \cite{Gurtin2010} and is well supported by experimental data \cite{Hill1950,Taylor1931,Chakrabarty2012}; therefore, it is used extensively in modern constitutive equations (e.g. \cite{Lele2009,Fleck2009,Zhu2016,Zhu2016a,Kundu2012}). As co-directionality is a sufficient condition for the alignment of principal directions, all materials consistent with co-directionality will be consistent with the alignment assumption.

Force balance and the alignment assumption are the only assumptions used, hence, the derived system of equations will be valid for a range of deformation regimes and material classes, regardless of the geometry and loading condition. These include many metals, ceramics, granular materials, elastomers, glasses, polymers and fluids. The set of assumptions made will be valid in a number of cases that cannot be analyzed using conventional experimental methodologies. In particular, in cases where heterogeneous stress fields arise that are not possible to measure, such as those caused by strain softening, high strain rates, material heterogeneity, complex geometries and complex loading conditions. See \S \ref{sec:range_of_applicability} for a more extended discussion on the applicability of the approach. 

\subsection{Infinitesimal strain}
\label{sec:derivation}
Here we derive equations for infinitesimal strain plasticity and later show how the results can be generalized to finite strain elasticity and plasticity (\S \ref{sec:elasto_visco_plastic} and \ref{sec:finite_strain}).
In this case we assume the principal directions of the Cauchy stress $\mathbf{T}$ are aligned with the strain rate $\mathbf{\dot{E}}$. Note that for this to be valid one must assume that $\mathbf{\dot{E}^e}<<\mathbf{\dot{E}^p}$ and hence that  $\mathbf{\dot{E}}\approx \mathbf{\dot{E}^p}$ \footnote{Here, $\mathbf{E^e}$ and $\mathbf{E^p}$ correspond to the elastic and plastic parts of the small strain such that $\mathbf{E} = \mathbf{E^e}+\mathbf{E^p}$.}. Applying the spectral theorem to $\mathbf{\dot{E}}$ and $\mathbf{T}$ we have\footnote{Notation: We use bold upper-case letters to denote second order tensors ($\mathbf{A},\mathbf{B}$,...). We denote the deviatoric part of a second order tensor $\mathbf{A}$ as $\mathbf{A_0}$, and the material time derivative as $\mathbf{\dot{A}}$ or $\mathbf{\dot{\overline{A}}}$. $\mathrm{Div}\mathbf{A}$ and $\mathrm{div}\mathbf{A}$ correspond to the divergence $\mathbf{A}$ with respect to the initial and deformed reference frame respectively. Similarly, $\mathrm{Grad}\mathbf{A}$ and $\mathrm{grad}\mathbf{A}$ correspond to the gradient of $\mathbf{A}$ with respect to the initial and deformed reference frame respectively. $|\mathbf{A}|=\sqrt{A_{ij}A_{ij}}$ gives the magnitude of $\mathbf{A}$. The spectral decomposition of a symmetric tensor $\mathbf{A}$ is written as $\mathbf{A} = \mathbf{Q_A\Lambda_AQ_A^T}$ where $\mathbf{Q_A}$ has columns with the Eigen vectors (principal directions) and $\mathbf{\Lambda_A}$ is a diagonal matrix containing the Eigen values (principal values). We use circular parentheses $(\cdot )$ to denote functional dependence and square brackets $[\cdot ]$ to specify the order of operands.}

\begin{equation}
    \mathbf{\dot{E}} = \mathbf{Q_{\dot{E}}\Lambda_{\dot{E}} Q^T_{\dot{E}}}, \quad \mathbf{T} = \mathbf{Q_T\Lambda_T Q^T_T}.
\end{equation}

Because the principal axis are aligned we have
\begin{equation}
    \mathbf{Q_T} = \mathbf{Q_{\dot{E}}}.
\end{equation}
Hence,
\begin{equation}
    \mathbf{T} = \mathbf{Q_{\dot{E}}\Lambda_T Q^T_{\dot{E}}}.
\end{equation}
Since $\mathbf{Q_{\dot{E}}^TQ_{\dot{E}}=I}$ we have
\begin{equation*}
    \mathbf{\Lambda_T} = \mathbf{Q^T_{\dot{E}}Q_{\dot{E}}\Lambda_TQ^T_{\dot{E}}Q_{\dot{E}}} =\mathbf{Q^T_{\dot{E}}TQ_{\dot{E}}}.
\end{equation*}
Therefore, the symmetric matrix $\mathbf{Q^T_{\dot{E}}TQ_{\dot{E}}}$ must be diagonal, leading to three equations specifying that each of the off diagonal terms are zero. Since $\mathbf{Q_{\dot{E}}}$ is computed as a function of $\mathbf{\dot{E}}$, these equations are linear in the coefficients of $\mathbf{T}$. Note that geometrically, we have just rotated the $\mathbf{T}$ matrix so that the basis vectors are the principal directions of $\mathbf{\dot{E}}$, then we specified that the shear components are zero.  In addition to this requirement, force balance must also hold giving rise to six total equations

\begin{equation}
    [\mathbf{Q^T_{\dot{E}}TQ_{\dot{E}}}]_{ij} = 0 \qquad for\quad i\neq j,
     \label{eq:codir3}
\end{equation}
\begin{equation}
    \mathrm{Div}{\mathbf{T}} + \mathbf{b} = \rho\dot{ \mathbf{v}},
    \label{eq:force3}
\end{equation}

where $\mathbf{b}$ is the body force, $\rho$ is the density and $\mathbf{v}$ is the velocity of a material point. Since the deformation field is known (i.e. $\mathbf{Q_{\dot{E}}}$ and $\mathbf{v}$), Eqs. (\ref{eq:codir3}-\ref{eq:force3}) give six equations with six unknowns and form a mathematically determinate, hyperbolic, first order, linear system of partial differential equations with variable coefficients. In Cartesian coordinates, Eq. (\ref{eq:codir3}) can be written as

\begin{equation}
    c_{1}T_{xx} + c_{2}T_{xy} + c_{3}T_{xz} + c_{4}T_{yy}+ c_{5}T_{yz} + c_{6}T_{zz}, = 0\label{eq:codir3_1}
\end{equation}
\begin{equation}
    c_{7}T_{xx} + c_{8}T_{xy} + c_{9}T_{xz} + c_{10}T_{yy}+ c_{11}T_{yz} + c_{12}T_{zz} = 0, \label{eq:codir3_2}
\end{equation}
\begin{equation}
    c_{13}T_{xx} + c_{14}T_{xy} + c_{15}T_{xz} + c_{16}T_{yy}+ c_{17}T_{yz} + c_{18}T_{zz} = 0, \label{eq:codir3_3}
\end{equation}
where each of these coefficients $c_i$ are functions of $\mathbf{\dot{E}}$ and have analytical expressions using the $\dot{E}_{ij}$ components. In addition, Eq. (\ref{eq:force3}) can be written as
\begin{equation}
    \frac{\partial T_{xx}}{\partial x} + \frac{\partial T_{xy}}{\partial y} + \frac{\partial T_{xz}}{\partial z} + b_x = \rho\dot{v}_x, \label{eq:force3_1}
\end{equation}
\begin{equation}
    \frac{\partial T_{xy}}{\partial x} + \frac{\partial T_{yy}}{\partial y} + \frac{\partial T_{yz}}{\partial z} + b_y = \rho\dot{v}_y, \label{eq:force3_2}
\end{equation}
\begin{equation}
    \frac{\partial T_{xz}}{\partial x} + \frac{\partial T_{yz}}{\partial y} + \frac{\partial T_{zz}}{\partial z} + b_z = \rho\dot{v}_z. \label{eq:force3_3}
\end{equation}

As this system of equations is hyperbolic, the boundary conditions necessary depend on the characteristic lines, which in turn depend on the coefficients $c_i$. These are discussed in \S \ref{sec:stress_propagation}.

We make some remarks:
\begin{itemize}
    \item In the case of quasi-static plastic deformation where $\mathbf{\dot{E}}\rightarrow0$, $\mathbf{\dot{E}}$ can simply be replaced with the strain increment $d\mathbf{E}$.
    \item In the case of viscous deformation we also know that the principal directions of $\mathbf{T}$ and $\mathbf{\dot{E}}$ are aligned. Hence, Eq. (\ref{eq:codir3}) applies in the small strain case. The expressions for finite strain are given in \S \ref{sec:finite_strain}.
    \item In the case where two or three of the principal values of $\mathbf{\dot{E}}$ are the same, any perpendicular principal directions consistent with the spectral decomposition can be chosen to give a valid solution.
    \item If $\rho$ is also unknown the continuity equation (conservation of mass) can be used in conjunction with the measured deformation field to compute $\rho$ from an initial condition. Hence, we still have a complete set of equations.
\end{itemize}

\subsection{Elastic deformation}
\label{sec:elasto_visco_plastic}
Equations of the same form can be derived for isotropic elastic deformation. Previously we assumed alignment of the principal directions of $\mathbf{T}$ and $\mathbf{\dot{E}}$. Instead of this, for isotropic elasticity we have that the principal directions of $\mathbf{T}$ are aligned with the principal directions of $\mathbf{E}$ \cite{Gurtin2010}. Hence, the same argument in \S \ref{sec:derivation} holds where $\mathbf{\dot{E}}$ is replaced with $\mathbf{E}$. Therefore, we have

\begin{equation}
    [\mathbf{Q^T_ETQ_E}]_{ij} = 0 \qquad \mathrm{for}\quad i\neq j,
    \label{eq:codir_small_elastic}
\end{equation}
where the spectral decomposition of $\mathbf{E}$ is given by
\begin{equation}
    \mathbf{E} = \mathbf{Q_E\Lambda_EQ_E^T}.
\end{equation}

In general the principal directions of $\mathbf{\dot{E}}$ and $\mathbf{E}$ are different, so one cannot directly apply our approach to elasto-plastic deformation. In this case, the decomposition of $\mathbf{E}$ and $\mathbf{\dot{E}}$ into its elastic and plastic components will be unknown. However, there are cases where the principal directions of $\mathbf{E}$ and $\mathbf{\dot{E}}$ are aligned throughout deformation and Eq. (\ref{eq:codir_small_elastic}) can also be applied (see \S \ref{sec:allignment_violation}).

\subsection{Generalization to finite strain}
\label{sec:finite_strain}
We use standard notation to specify the kinematics of large deformation. We have the position of a point in the reference body $\mathbf{X}$ and a smooth one to one mapping to the deformed body $\mathbf{x}=\boldsymbol{\chi}(\mathbf{X},t)$. We then define the deformation gradient, velocity, and velocity gradient by

\begin{equation}
    \mathbf{F} = \mathrm{Grad}\boldsymbol{\chi}, \qquad \mathbf{v} = \dot{\boldsymbol{\chi}}, \qquad \mathbf{L} = \mathrm{grad}\mathbf{v} = \dot{\mathbf{F}}\mathbf{F^{-1}}.
\end{equation}
Applying a polar decomposition to $\mathbf{F}$ gives

\begin{equation}
    \mathbf{F} = \mathbf{RU}=\mathbf{VR},
\end{equation}
where 
\begin{equation}
    \mathbf{U} = \sqrt{\mathbf{F^TF}}, \qquad \mathbf{V} = \sqrt{\mathbf{FF^T}},
\end{equation}
are the right and left stretch tensors respectively. The right and left Cauchy-Green deformation tensors are

\begin{equation}
    \mathbf{C} = \mathbf{U^2} = \mathbf{F^TF}, \qquad \mathbf{B} = \mathbf{V^2} =  \mathbf{FF^T}.
\end{equation}

$\mathbf{L}$ can be decomposed into its symmetric and skew parts

\begin{equation}
    \mathbf{D} = \frac{1}{2}[\mathbf{L}+\mathbf{L^T}], \qquad
    \mathbf{W} = \frac{1}{2}[\mathbf{L}-\mathbf{L^T}],
\end{equation}
which are known as the stretching and spin tensors respectively.

For the case of plastic, viscous, or visco-plastic deformation, $\mathbf{T}$ will have its principal axis aligned with the stretching tensor $\mathbf{D}$ \cite{Lele2009}. For the case of elastic deformation, $\mathbf{T}$ will have its principal axes aligned with $\mathbf{B}$, see for example, a Moony Rivlin material \cite{Mooney1940, Rivlin1948}. Hence, for large strains we have

\begin{equation}
    \mathrm{div}\mathbf{T}+\mathbf{b} = \rho\dot{ \mathbf{v}},
    \label{eq:force_finite_strain}
\end{equation}
\begin{equation}
    \begin{aligned} 
      [\mathbf{Q^T_DTQ_D}]_{ij} = 0 \quad \textrm{for}\quad i\neq j \qquad & \textrm{visco-plastic deformation}, \\ 
      [\mathbf{Q^T_BTQ_B}]_{ij} = 0 \quad \textrm{for}\quad i\neq j \qquad & \textrm{elastic deformation},
   \end{aligned}
   \label{eq:codir_finite_strain}
\end{equation}

where the spectral decompositions of $\mathbf{D}$ and $\mathbf{B}$ are
\begin{equation}
    \mathbf{D} = \mathbf{Q_D\Lambda_DQ_D^T}, \qquad \mathbf{B} = \mathbf{Q_B\Lambda_BQ_B^T}.
\end{equation}

\subsection{Two dimensional formulation}
\label{sec:two_dimensional_formulation}
The two-dimensional forms of Eqs. (\ref{eq:codir3}-\ref{eq:force3}) have applicability in several areas, for example, when the strain is measured on the surface of a sample using digital image correlation. In two dimensional quasi-static conditions with no body force, Eqs. (\ref{eq:codir3_1}-\ref{eq:force3_3}) reduce to 

\begin{equation}
\label{eq:forcex}
    \frac{\partial T_{xx}}{\partial x} + \frac{\partial T_{xy}}{\partial y} = 0,
\end{equation}

\begin{equation}
\label{eq:forcey}
    \frac{\partial T_{xy}}{\partial x} + \frac{\partial T_{yy}}{\partial y} = 0,
\end{equation}

\begin{equation}
    T_{xx}-T_{yy}-cT_{xy} = 0,
    \label{eq:codir2}
\end{equation}
with 
\begin{equation}
    c = \frac{\dot{E}_{xx}-\dot{E}_{yy}}{\dot{E}_{xy}}.
    \label{eq:c}
\end{equation}
These equations hold for plane stress, plans strain, or any other stress state with $T_{xz}=T_{yz}=E_{xz}=E_{yz}=0$. This gives three equations and three unknowns.

Here, we can define the Airy stress function $\psi$ such that
\begin{equation}
    T_{xx} = \frac{\partial \psi}{\partial y^2}, \quad
    T_{yy} = \frac{\partial \psi}{\partial x^2}, \quad
    T_{xy} = -\frac{\partial \psi}{\partial x \partial y}.
    \label{eq:airy_def}
\end{equation}

When this is substituted into Eqs. (\ref{eq:forcex}-\ref{eq:forcey}), the equations are automatically satisfied due the equality of mixed partial derivatives. We are left with one governing equation arising from Eq. (\ref{eq:codir2}):

\begin{equation}
\label{eq:codir_airy}
     \frac{\partial \psi}{\partial y^2} -  \frac{\partial \psi}{\partial x^2} + c\frac{\partial \psi}{\partial x \partial y} = 0.
\end{equation}

Note that $c$ specifies the angle of the principal directions with respect to the horizontal with $c=0$ corresponding to $\theta = \pi/4$, $c \rightarrow{}\infty$ corresponding to $\theta = 0$ and $c \rightarrow{}-\infty$ corresponding to $\theta = \pi/2$. More generally,

\begin{equation}
\label{eq:theta}
    \theta = tan^{-1}\left(\frac{1}{c+\sqrt{1+c^2}}\right).
\end{equation}

Also note that when $c=0$ everywhere, Eq. (\ref{eq:codir_airy}) reduces to the classic second order wave equation.

\section{Numerical proof of principle}
\label{sec:proof_of_principle}

\begin{figure}[ht!]
\centering
\includegraphics[width=0.8\textwidth]{{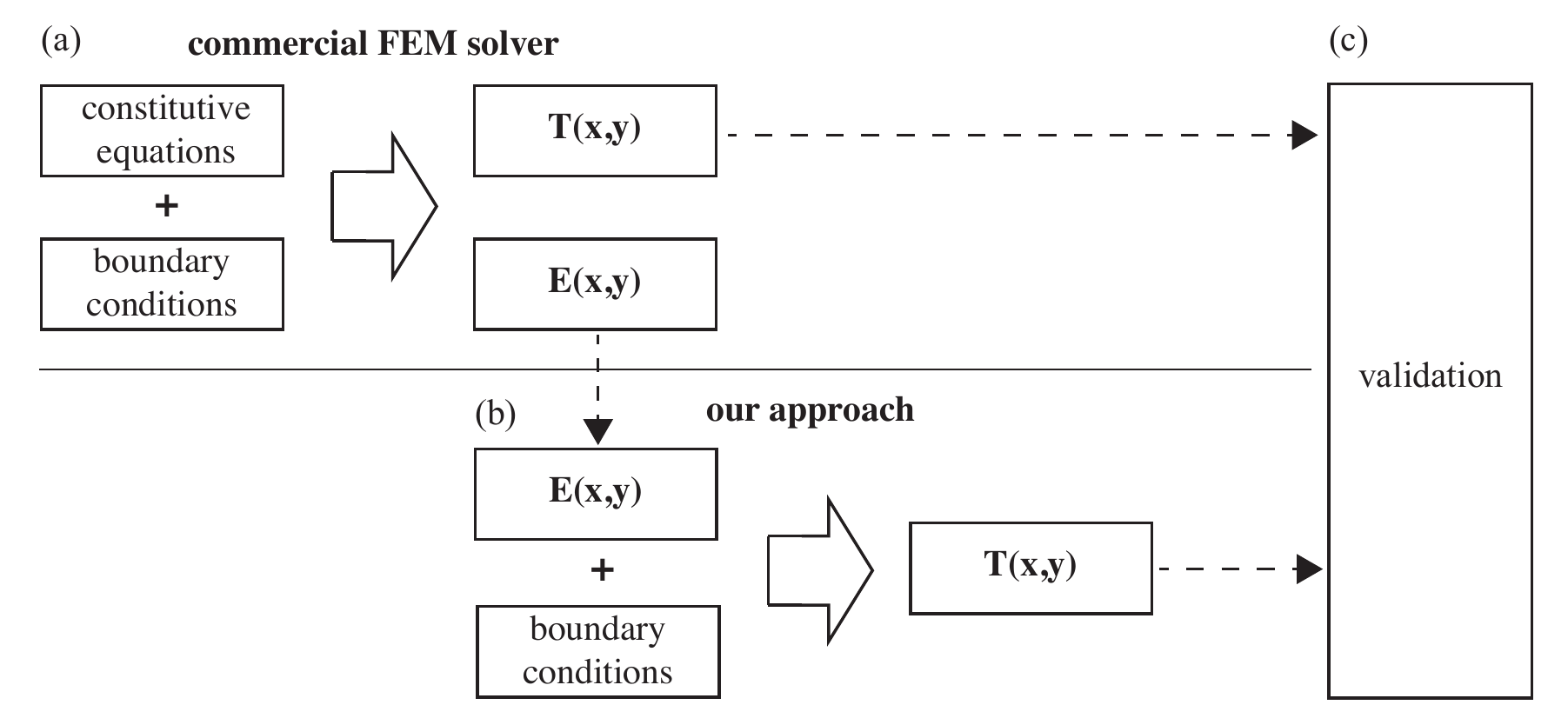}}
\caption{The procedure used to numerically validate and test our approach. (a) The problem is specified with spatially varying constitutive equations and boundary conditions (see Fig. \ref{fig:inverse}) and solved using a commercial FEM solver. (b) The strain data output from (a) is used with the boundary conditions to compute the stress using our approach. (c) The stresses computed from (a) and (b) are compared to validate the approach. }\label{fig:methodology}
\end{figure}

We consider the elasto-plastic deformation of a bar in plane stress where the yield stress and hardening behavior vary as a function of position (shown in Fig. \ref{fig:inverse}a). The objective is to show that we can accurately compute the stress without any knowledge of of constitutive equations and how they vary; only knowledge of the strain field and boundary conditions is used. Fig. \ref{fig:methodology} summarizes the methodology used to validate our approach. First, the constitutive equations and boundary conditions are input into a commercial finite element solver which then outputs stress and strain fields (Fig. \ref{fig:methodology}a). Second, the strain field and boundary conditions are used with our equations in order to compute a new stress field (Fig. \ref{fig:methodology}b). Third, this stress field is compared to the original stress field and the error is evaluated (Fig. \ref{fig:methodology}c). The example was inspired by common metallurgical tension experiments where the boundary conditions can be estimated from the load cell, the strain field can be measured using digital image correlation, but the local stress field becomes unknown when necking occurs (e.g. \cite{Zhang2017}).

\subsection{Forward problem}
\label{sec:numerical_problem}
A two-dimensional rectangular bar with initial dimensions of $l_x$ in the $x$ direction and $l_y=l_x/4$ in the $y$ direction is subject to a constant traction perpendicular to its boundary at each end. The material is elasto-plastic with a Mises yield surface and co-directionality assumed. The material has an elastic modulus $E = 200$ GPa and a Poisson's ratio $\nu = 0.3$. The material has linear hardening with a hardening coefficient $H$ and initial yield stress $Y$ that both vary as a function of position (these are specified in Fig. \ref{fig:inverse}a). The yield stress varies between 50 and 100 MPa, and the hardening coefficient takes values between 3 and 4 GPa. There is no kinematic hardening. The computation was done for one elasto-plastic stress increment with a boundary stress of $\sigma=120$ MPa applied at the boundary. Here all the material is yields. Specifically, the boundary conditions are

\begin{equation}
    \begin{split}
        T_{xx}(x=0,y) = \sigma, \qquad
        T_{xx}(x=l_x,y) = \sigma,\\
        T_{xy}(x=0,y) = 0,\qquad
        T_{xy}(x=l_x,y) = 0,\\
        T_{xy}(x,y=0) = 0,\qquad
        T_{xy}(x,y=l_y) = 0,\\
        T_{yy}(x,y=0) = 0,\qquad
        T_{yy}(x,y=l_y) = 0.\\
    \end{split}
    \label{eq:boundary_full}
\end{equation}

\begin{figure}[ht!]
\centering
\includegraphics[width=1\textwidth]{{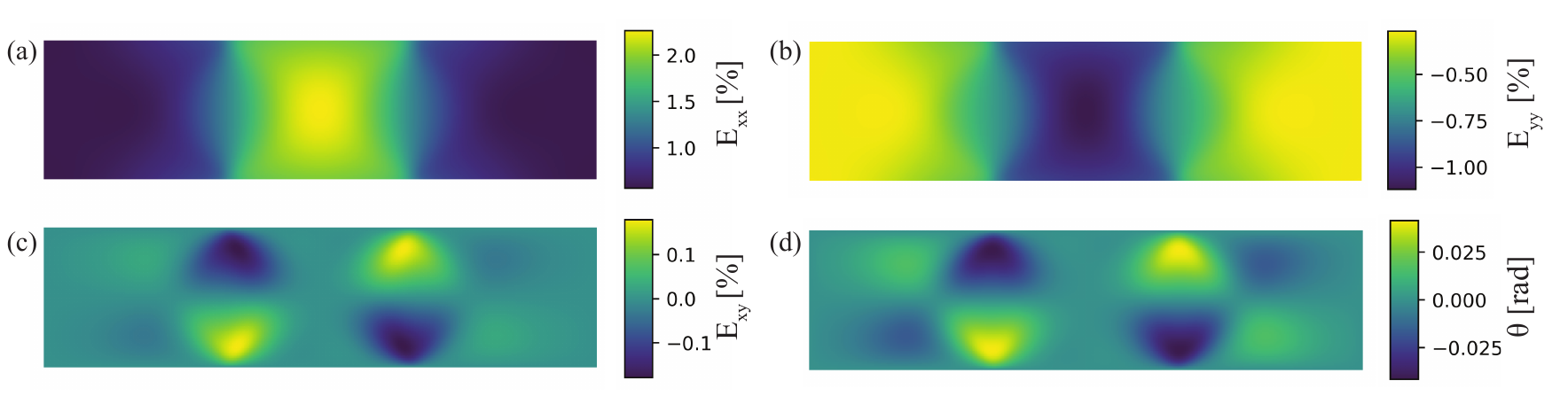}}
\caption{(a-c) The strain data output from the commercial finite element model. These are used to compute (d) the angle of the principal directions of stress $\theta$. These principal directions are used in our approach to compute the stress. Note that there are minor differences in (c) and (d) even though they appear similar. }\label{fig:strain}
\end{figure}

The stress and strain fields were computed using a commercial finite element solver ABAQUS/standard (2017). There were 100 equally sized cuboid elements in the $y$ direction, 400 elements in the $x$ direction, and 1 element in the $z$ direction, giving 40000 total elements. The elements were 20 node bricks, triquadratic displacement, hybrid, linear pressure and reduced integration \cite{abaqus2017}.  The thickness of the bar was specified to be $l_x/1000$ to ensure plane stress conditions. The strain output is shown in Fig. \ref{fig:strain}(a-c) and is used as input into the inverse problem approach. 

\subsection{Inverse problem}
\label{sec:numerical_implementation}
\subsubsection{Problem specification}
As Eq. (\ref{eq:codir_airy}) is hyperbolic, the boundary conditions previously given over specify the problem. Instead we use

\begin{equation}
    \begin{split}
        T_{xx}(x=0,y) = \sigma, \qquad
        T_{xx}(x=l_x,y) = \sigma,\\
        T_{xy}(x,y=0) = 0, \qquad
        T_{yy}(x,y=0) = 0.\\
    \end{split}
    \label{eq:bc_stress2}
\end{equation}
though there are other subsets of Eq. (\ref{eq:boundary_full}) which also would have been sufficient (\S \ref{sec:stress_propagation}). Note that the other boundary condition values are computed as part of the solution. The stress Airy function $\psi$ is specified up to a plane, i.e. for any valid solution $\psi$ we have another valid solution $\psi'$ given by
\begin{equation}
    \psi' = \psi + a_1 + a_2x + a_3y,
    \label{eq:airy_plane}
\end{equation}

for any  $a_1$, $a_2$ and $a_3$. In order to remove these degrees of freedom we fix the following values

\begin{equation}
    \psi(x=0,y=0)=0, \quad
    \psi(x=l_x,y=0)=0, \quad
    \psi(x=0,y=l_y)=0.
    \label{eq:airyfix}
\end{equation}

From Eqs. (\ref{eq:airy_def}) and (\ref{eq:bc_stress2}) we have
\begin{equation}
    \begin{split}
        \frac{\partial^2\psi}{\partial y^2}(x=0,y)=\sigma,\\ 
        \frac{\partial^2\psi}{\partial x\partial y}(x,y=0)=0,\\
        \frac{\partial^2\psi}{\partial x^2}(x,y=0)=0,\\
        \frac{\partial^2\psi}{\partial y^2}(x=l_x,y)=\sigma.
    \end{split}
    \label{eq:bc_airy1}
\end{equation}
However, we cannot directly apply these boundary conditions as they are second order in the derivatives of $\psi$. Nevertheless, we can combine Eq. (\ref{eq:bc_airy1}) with Eq. (\ref{eq:airyfix}) to give usable boundary conditions. Our problem now becomes
\begin{equation}
    \begin{split}
        \textrm{solve:}&\\
        &\frac{\partial \psi}{\partial y^2} -  \frac{\partial \psi}{\partial x^2} + c\frac{\partial \psi}{\partial x \partial y} = 0,\\
        \textrm{subject to:}&\\
        &\psi(x=0,y) =\sigma [y^2-l_yy]/2,\\
        &\psi(x=l_x,y) =\sigma [y^2-l_yy]/2,\\
        &\psi(x,y=0) = 0,\\
        &\frac{\partial\psi}{\partial y}(x,y=0) =-\sigma l_y/2,
    \end{split}
    \label{eq:numerical_problem}
\end{equation}
where $c$ as a function of $x$ and $y$ is given. As the stress in computed in the forward problem over just one increment, the principal directions of the strain and strain rate will be aligned, hence one can just compute $c$ using the strain field. Specifically, $c$ is computed from the strain values output from the forward problem (Fig. \ref{fig:strain}a-c) using Eq. (\ref{eq:c}) (Fig. \ref{fig:strain}d). 

\subsubsection{Numerical implementation}
\label{ssec:numerical_implementation}
Although the governing equations are linear, there are several challenges in developing a robust numerical method due to the variable coefficient $c$ and the susceptibility of hyperbolic numerical methods to be becoming unstable. The system of governing equations has no preferred direction of the characteristic curves complicating the problem further (unlike many hyperbolic systems where characteristic curves go forward in time).  Here, the objective it simply to  demonstrate the validity of the governing equations and overall approach. Hence, we implement a simplified finite difference based approach that exploits the fact that the characteristic lines deviate from the $x$ and $y$ axis by only a small amount, and that the geometry is simple. However, development of more robust numerical methods, such as those using the finite volume method, is likely warranted.

\begin{figure}[ht!]
\centering
\includegraphics[width=0.6\textwidth]{{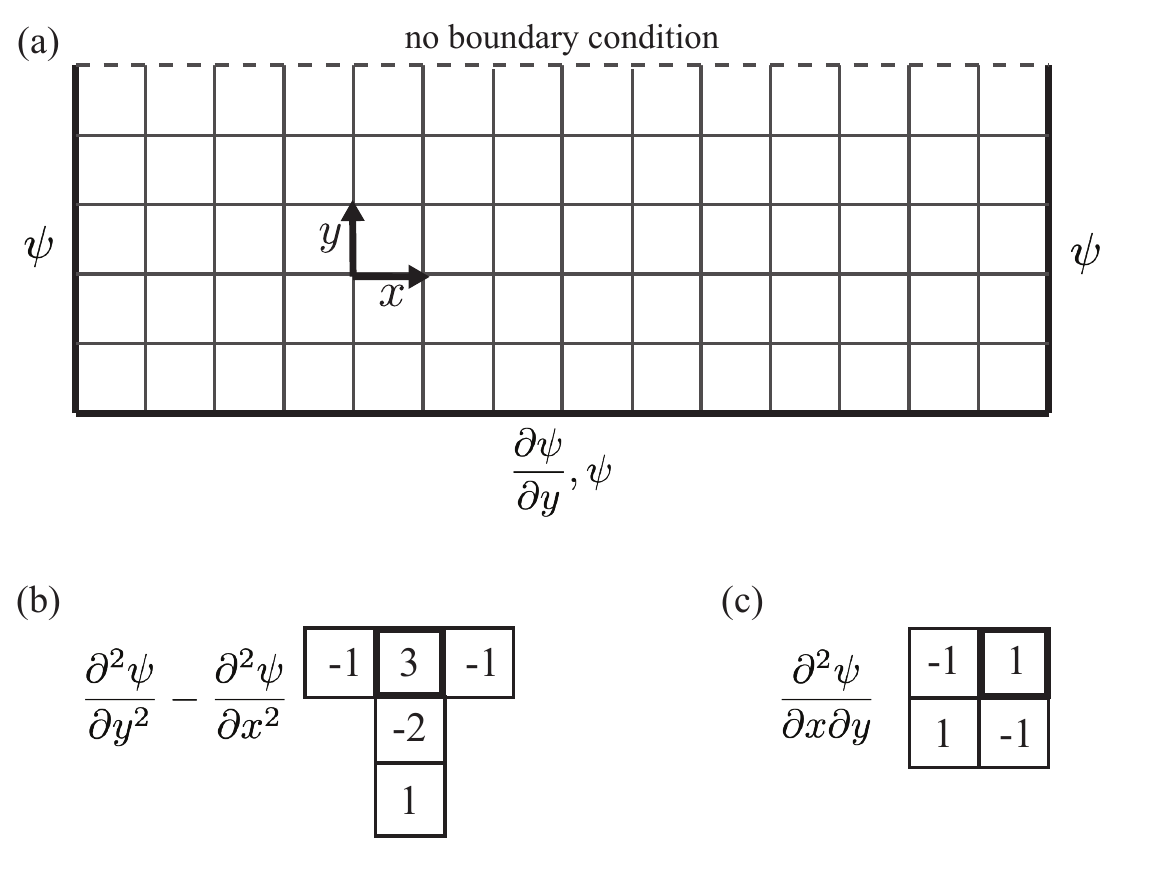}}
\caption{The finite difference discretization approach used for the computation of our solution. (a) The domain is divided into a rectangular mesh with three boundary conditions specifying $\psi$ and one specifying $\frac{\partial \psi}{\partial y}$. Note that there are more mesh points used in the actual solution. (b-c) The finite difference stencils used to approximate the  $\psi$ derivatives. }\label{fig:discretization}
\end{figure}

We implement the solution using a finite difference discretization. As the equation is hyperbolic, we use methods analogous to those that are implicit with respect to time. A rectangular mesh is used with 100 by 400 nodes and the boundaries are specified on the right, left and bottom sides (Fig. \ref{fig:discretization}a). The stencils used to approximate the various $\psi$ derivatives from Eq. (\ref{eq:numerical_problem}) are shown in Fig. \ref{fig:discretization}b-c. See the Appendix for further details on the discretization and its implementation. 

Solving the system took approximately 0.4 seconds (on a 2014 Macbook Pro with a 2.5 GHz Intel Core i7 processor and 8 GB RAM), and future speedups could could be achieved by exploiting the structure of the matrix, for example, by using fast Poisson solvers \cite{Strang2007}. To put this in perspective, note that the commercial finite element solver, ABAQUS/standard (2017) took 23 minutes to simulate the stress and strain (on a windows desktop computer with an 3.2 GHz Intel Cor i5 processor and 8 GB RAM). This difference in computational time of a factor of several thousand is significant and highlights the fact that Eq. (\ref{eq:numerical_problem}) is linear, as opposed to the non-linear equations of the forward problem.

Here we make two remarks:
\begin{itemize}
    \item It was not necessary to express the problem using $\psi$, one could have used Eqs. (\ref{eq:forcex}-\ref{eq:codir2}), and Eq. (\ref{eq:bc_stress2}).
    \item It was necessary to remove a subset of the boundary conditions in Eq. (\ref{eq:boundary_full}). Not doing so introduces numerical instabilities known as parasite modes \cite{Trefethen1982}. Furthermore, given the boundary conditions chosen, it was necessary to choose a backward difference scheme to approximate $\frac{\partial^2 \psi}{\partial y^2}$, and $\frac{\partial^2 \psi}{\partial x \partial y}$.
\end{itemize}

\subsection{Validation of solution}
\label{sec:validation}
\begin{figure}[ht!]
\centering
\includegraphics[width=1\textwidth]{{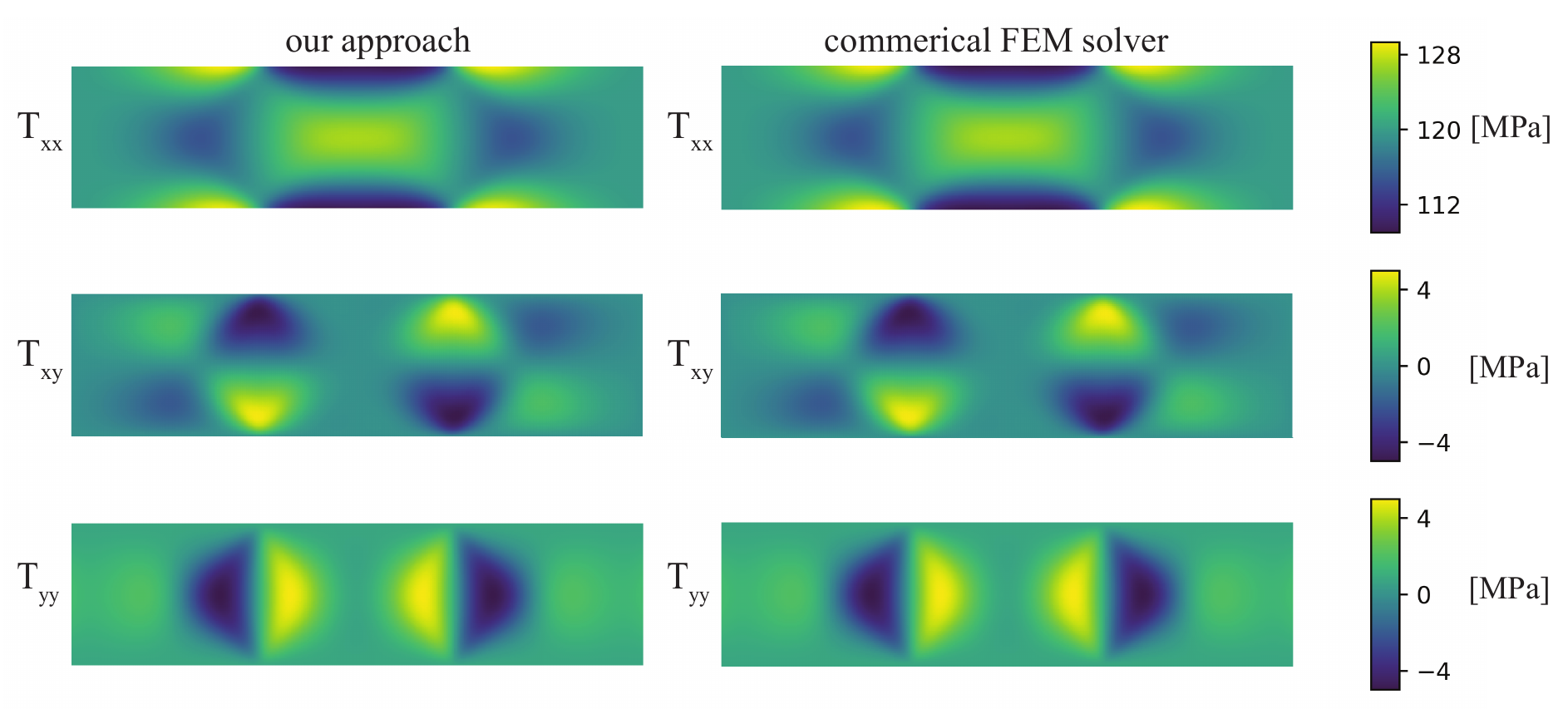}}
\caption{A comparison of the stress computed for the inverse problem with our approach and the stress computed for the forward problem by the commercial finite element solver.}\label{fig:validation_field}
\end{figure}

\begin{figure}[ht!]
\centering
\includegraphics[width=1\textwidth]{{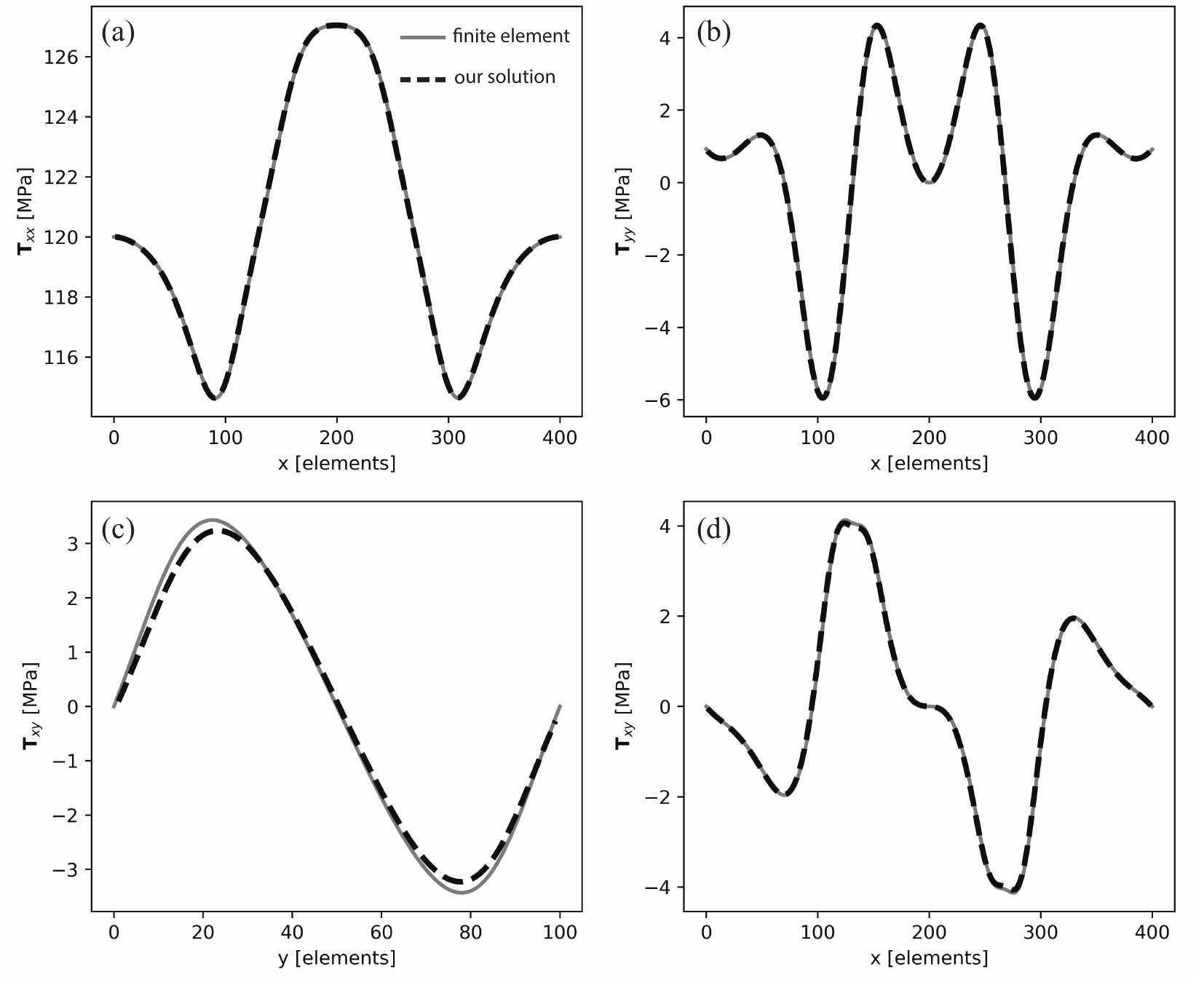}}
\caption{A comparison of the stress computed for the inverse problem with our approach and the stress computed for the forward problem by the commercial finite element solver. (a) $T_{xx}$ plotted as a function of $x$ with $y=l_y/2$. (b) $T_{yy}$ plotted as a function of $x$ with $y=l_y/2$. (c) $T_{xy}$ plotted as a function of $y$ with $x=3l_x/4$. (d) $T_{xy}$ plotted as a function of $x$ with $y=l_y/4$.}\label{fig:validation_line}
\end{figure}

The  solution produced is close to exact over the entire domain, this is shown in Figs. \ref{fig:validation_field} and \ref{fig:validation_line}. The normalized root mean squared error (NRMSE), normalized by the applied stress $\sigma$, is computed using:

\begin{equation}
    \mathrm{NRMSE}(\mathbf{T}) = 
    \frac{\sqrt{\mathrm{mean}(|\Delta\mathbf{T}|^2)}}{\sigma},
\end{equation}

where $\Delta \mathbf{T}$ is the difference between the stress computed from the inverse problem and the stress output by the commercial finite element solver the mean is taken over all positions except nodes on the boundary which are known from the boundary conditions\footnote{Note that the error on the boundary nodes themselves was relatively high, however this problem should be addressed by developing more robust numerical methods.}. The NRMSE takes a value of $1.63\times10^{-4}$, indicating accurate results. Note that there does appear to be small numerical errors in some regions, e.g. near the left and right side of Fig. \ref{fig:validation_line}c, however, this can likely be addressed by developing more robust numerical methods, for example by using similar approaches to those described in described in \cite{LeVeque2002}.

\subsection{Sensitivity to error in input data}
The strain $\mathbf{E}$ used was imported from numerical finite element software, and hence, is only subject to small numerical errors. However, the error in experimentally measured strain is typically much higher \cite{Pan2009}. We conduct a preliminary investigation into the sensitivity of our approach to the typical errors one might encounter when attempting to compute stress from DIC data on a solid sample. We consider three types of error in the input data: (i) random, non-spatially correlated, error the displacement field data (from which the input strain field is derived); (ii) random, non-spatially correlated error in the assumed boundary conditions; (iii) and systematic error in the assumed boundary conditions. Non-spatially correlated displacement error typically arises from DIC due to the finite pixel size and other imaging aberrations \cite{Hild2006}. We consider uncorrelated noise in the displacement field rather than the strain field because the strain field is computed from the gradient of the displacement field by fitting the slope using linear regression \cite{Blaber2015a}. As we show, the error in $\mathbf{T}$ (and $\mathbf{E}$) is highly dependent upon the choice of how many points to include in this linear regression. In addition, we consider uncertainties in the boundary conditions. These could arise due to sample heterogeneities and imperfect sample preparation (random error), or by making an imperfect approximations such as uniform stress along the gauge section of a tensile specimen (systematic error). Note that we do not investigate systematic error in the displacement field, such as lens distortion error, due to the complexity of the analysis required and the fact that many of these errors can be addressed using sophisticated optical setups or through image post processing.

In order to investigate these errors we use a modified strain field or modified boundary conditions for the inverse problem computation (Fig. \ref{fig:methodology}b). We have the same original $\mathbf{E}$, boundary conditions and $\mathbf{T}$ for the forward problem computation, and validate against these. To investigate the effect of random, non-spatially correlated noise, random Gaussian noise with a specified standard deviation $\mathrm{std}(u_i')$ is added to each component of the displacement field $\mathbf{u}(\mathbf{x})$. $\mathbf{E}(\mathbf{x})$ is then computed using linear regression over a number of displacement values surrounding each point, the same approach as used standard DIC software packages such as ncorr \cite{Blaber2015a}. The parameter $l_{\mathbf{u}\rightarrow\mathbf{E}}$ gives the length scale over which the strain is computed. This parameter plays a key role in determining the error in $\mathbf{T}$. To investigate random noise in the boundary conditions, random Gaussian noise is added to each point on the boundary with a specified standard deviation $\mathrm{std}(T_{xx}')$. To investigate systematic error in the boundary condition, an error that is a linear function of $y$ is chosen. For more a more details on this methodology, see the Appendix. 

\begin{figure}[ht!]
\centering
\includegraphics[width=1\textwidth]{{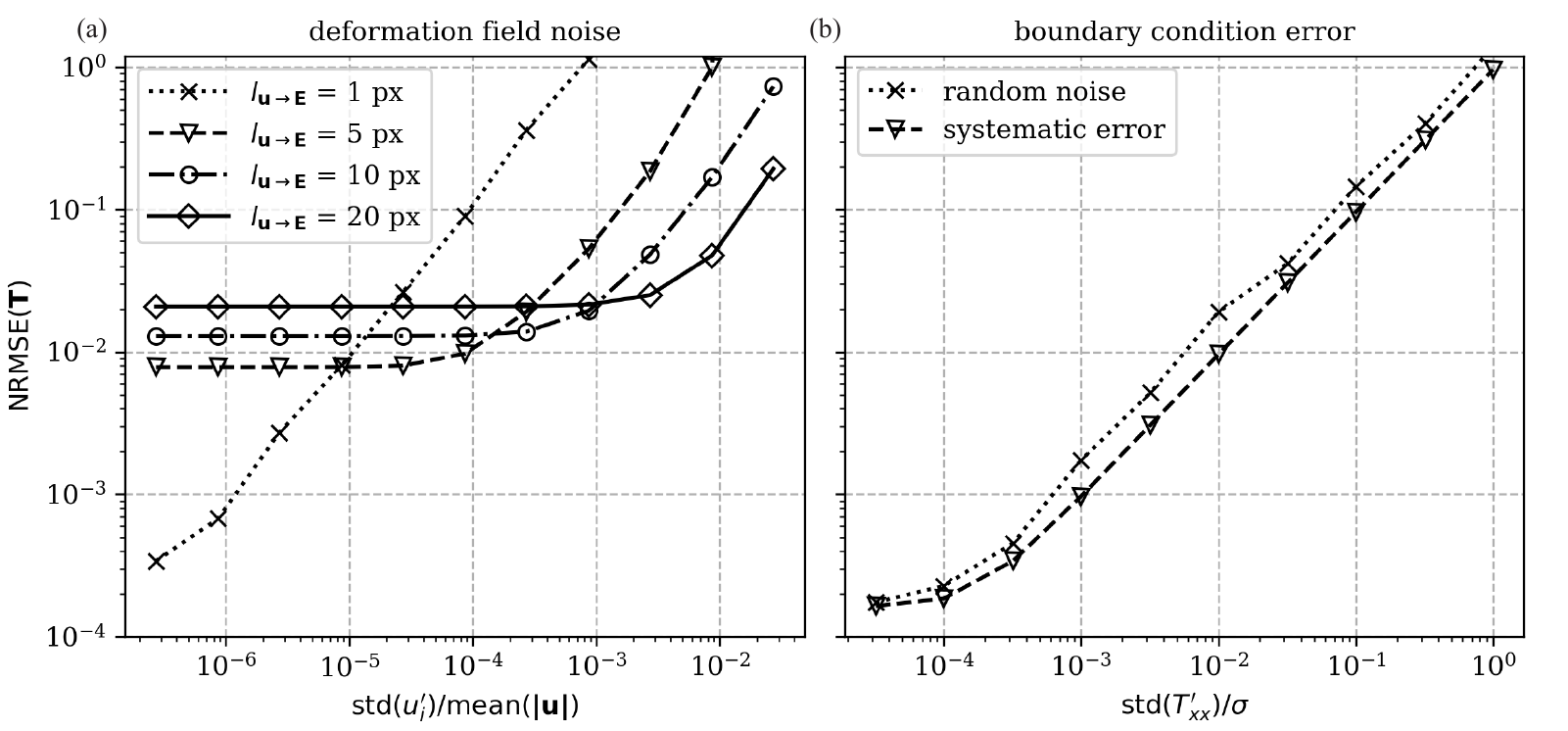}}
\caption{Effect of error in the input data on the error of the computed stress. (a) The NRMSE error in the stress as a function of standard deviation of the noise introduced into the displacement field. This depends strongly on the parameter $l_{\mathbf{u}\rightarrow\mathbf{E}}$ which controls the number of points used in the linear regression when computing $\mathbf{E}$. (b) Error in the computed stress as a function of error in the boundary condition for both random Gaussian noise and systematic error.}
\label{fig:noise_quantify}
\end{figure}

\begin{figure}[ht!]
\centering
\includegraphics[width=1\textwidth]{{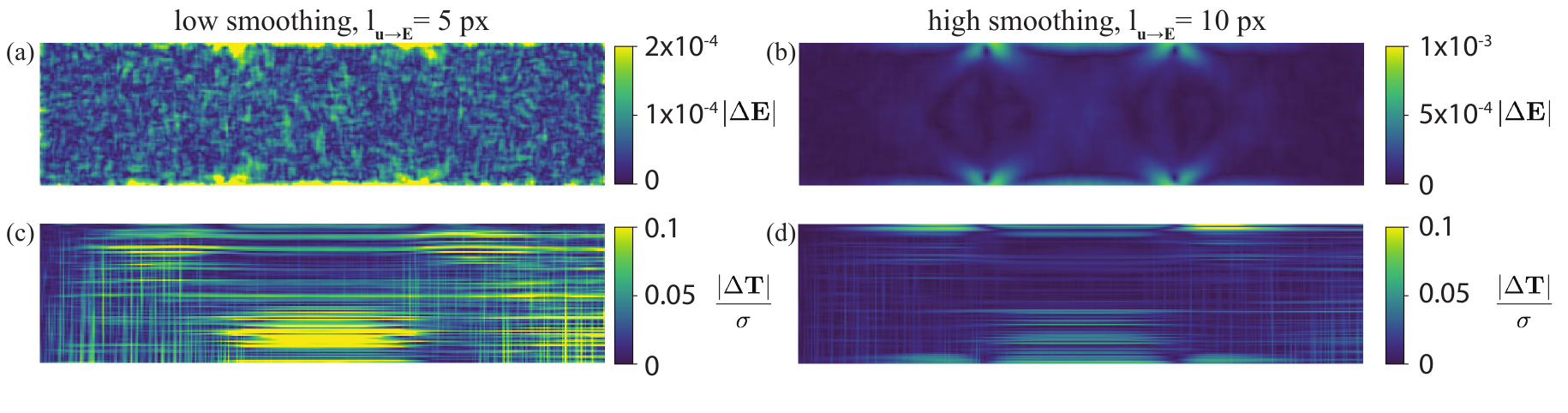}}
\caption{
Effect of error in the input data or the stress computed with changing $l_{\mathbf{u}\rightarrow\mathbf{E}}$. (a-b) The error in the strain. (c-d) The corresponding computed stress field. }\label{fig:noise_field}
\end{figure}

The mean error in stress is shown in Fig. \ref{fig:noise_quantify} and two examples of error as a function of position are shown in Fig. \ref{fig:noise_field}. Fig. \ref{fig:noise_quantify}a shows that the error in stress is highly dependent upon $l_{\mathbf{u}\rightarrow\mathbf{E}}$. There is a trade-off between using a short $l_{\mathbf{u}\rightarrow\mathbf{E}}$ which does not sufficiently filter out the noise (Fig. \ref{fig:noise_field}a), and a long $l_{\mathbf{u}\rightarrow\mathbf{E}}$ which does not sufficiently capture sharp gradients (Fig. \ref{fig:noise_field}b). Note that when computing $\mathbf{T}$ from noisy DIC data in practice, there are a variety of strategies one could use to select a $l_{\mathbf{u}\rightarrow\mathbf{E}}$ to minimize the error. These include assessing the sensitivity to additional added noise, using a comparison with similar numerical experiments, or simply making qualitative observations of the $\mathbf{T}(\mathbf{x})$ and $\mathbf{E}(\mathbf{x})$. Fig. \ref{fig:noise_quantify}b shows that error introduced due to an incorrect boundary condition is well approximated by the error in the boundary condition itself, up to errors that match the numerical method error with no input error. Furthermore, there is relatively little difference in random noise and systematic error. 

We remark that the error for this particular problem is likely to be acceptable for the typical noise values encountered in DIC. We have a $\mathrm{NRMSE}(\mathbf{T}) =$ 1.92 percent are obtained with $\mathrm{std}(u'_i)= 3.2\times10^{-6}l_x$, a root mean square error of the strain $\mathrm{RMSE}(\mathbf{E})=1.29\times 10^{-4}$, and $l_{\mathbf{u}\rightarrow\mathbf{E}}$ = 10 pixels (Fig. \ref{fig:noise_field}b,d). These are values commonly observed in typical DIC experiments \cite{Tong2005}. Note that $\mathrm{mean}(|\mathbf{u}|)=3.69\times 10^{-3}l_x$.  However, one should use caution when generalizing this error to other problems. As we discuss below, this error will be dependent on a number of problem parameters. We also remark that for this problem it would be experimentally challenging to reduce the error in the stress by an order of magnitude or more, as this would require several orders of magnitude in reduction of the displacement error.

In addition to the error of the displacement field, there are several factors which may impact the accuracy of $\mathbf{T}$. First, we consider the average magnitude of the strain. Typically, the error in the strain field will be approximately independent of average strain. Hence, the larger the average strain, the smaller the impact of noise on the principal directions of the strain from which $\mathbf{T}$ is computed. One can show that the error in the principal direction is inversely proportional to $|\mathbf{E}|$ if the error in $\mathbf{E}$ is much less $\mathbf{E}$ itself.  Hence, the algorithm will be more accurate at larger strains, such as those that arise during significant plastic deformation. Second, the magnitude of the strain gradients present in the deformation will impact the accuracy. Lower gradients will enable more accurate computation of $\mathbf{T}$ because more points can be used to compute the strain without overly smoothing the strain gradients. The problem investigated has relatively sharp gradients (Fig. \ref{fig:strain}a), so we would expect greater accuracy for a range of other problems. Third, we consider that the specific numerical algorithm used may contribute to the error, as numerical algorithms for hyperbolic systems can often be unstable. However, it is unclear how much error can be attributed to this or if the error is intrinsic in the partial differential equation with incorrect coefficients. We speculate that with development of more robust numerical algorithms it may be possible to obtain higher accuracy. Hence, the error in the stress will be dependent on the quality of the DIC data, the gradients in the strain, the magnitudes of the strain, and likely the numerical method used. Furthermore, there are likely to be several strategies one can use to reduce the error that the authors have not explored. These could include using higher order linear regression or alternate filtering approaches to capture sharper gradients while eliminating noise or adding regularization terms to the governing equations.

The error in $\mathbf{T}$ due to error in the boundary conditions is comparatively easier to understand. One can use linear superposition to split the stress field into the true solution with the correct boundary conditions, plus a stress error field due to the error in the boundary conditions. One can then use scaling analysis to estimate the magnitude of the stress error field. For the problem considered here, one can show that the magnitude of the stress directly scales with the magnitude of the stress at the boundary, hence the error in the stress scales with the error of the stress at the boundary. This explains the slope on Fig. \ref{fig:noise_quantify}b and why there is little difference between random noise and systematic error. We make one remark on the choice of the boundary in practice when given DIC data. One can use the alignment assumption with the experimental strain data to check the approximation used. For example, if one wants to check if the shear stress along the cross section of a tensile specimen is zero, one can check if the shear strain is zero along this boundary.

\section{Physical interpretation - stress propagation}
\label{sec:stress_propagation}
Since the governing equations are hyperbolic, it is mathematically sufficient to only specify a subset of the full traction boundary conditions \cite{Zachmanoglou1986}. However, it may be intuitively unclear why this is the case. Furthermore, it may be unclear how the stress can be computed everywhere without specifying the constitutive equations. Here we present a physical interpretation of how the solution is computed.

\begin{figure}[ht!]
\centering
\includegraphics[width=0.35\textwidth]{{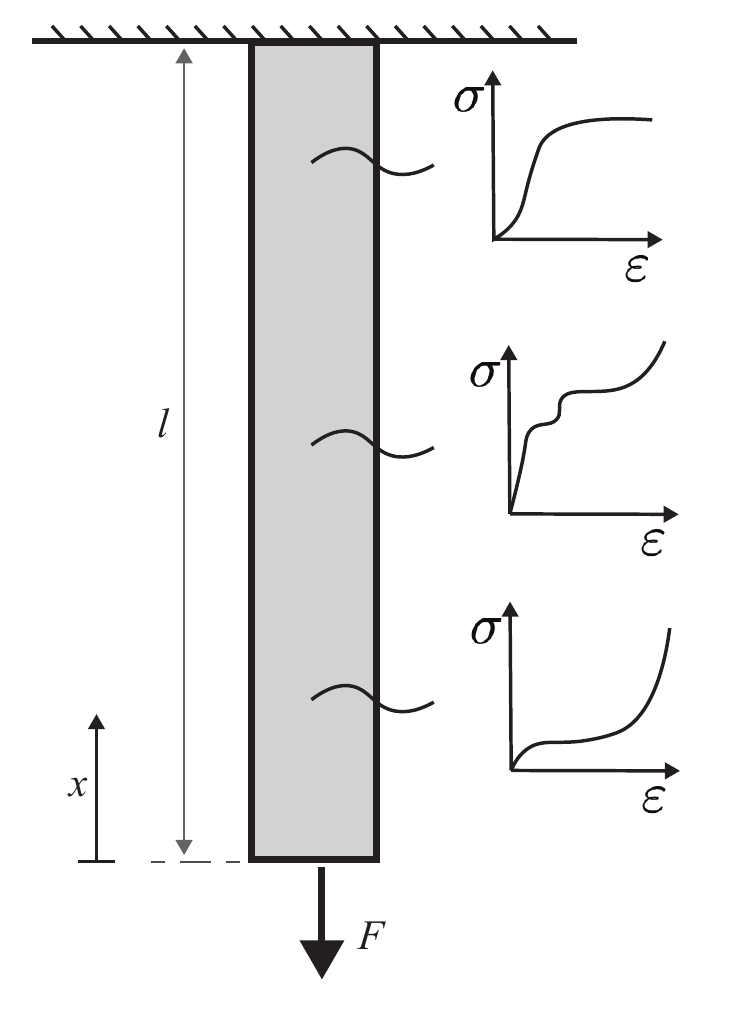}}
\caption{A one dimensional bar subject to an applied force $F$ at the boundary with constitutive equations that vary as a function of position. Here the force can be computed everywhere regardless of the complexity of these constitutive equations.}
\label{fig:one_dimension}
\end{figure}

We first consider the simplest case of static one-dimensional deformation, which shares key features with the multidimensional case. We consider a bar of constant cross section, deformed in tension, with a force at the boundary $F(x=0)=F$ (Fig. \ref{fig:one_dimension}). This has the trivial analytical solution of $F(x)=F$ everywhere, however, here we consider a discrete iterative solution with step size $\Delta x$. First, the force at $\Delta x$ must be $F$ in order to ensure that the material between $0$ and $\Delta x$ is at equilibrium. Second, the force at $2\Delta x$ must be $F$ in order to ensure that the material between $\Delta x$ and $2 \Delta x$ is at equilibrium. This logic can be applied to the whole bar giving rise to the solution $F(x)=F$ everywhere. This solution procedure corresponds to the force value at the boundary propagating through the material. Because of this, only a subset of all boundary conditions (one side) is required. Also note that this solution is valid despite the fact that no constitutive equations are specified; these could be very complex and vary as a function of position along the bar (Fig. \ref{fig:one_dimension}). The primary difference between this and the two and three dimensional cases is that there are multiple directions stress could propagate throughout the material, hence, we require the additional assumption specifying the principal directions of $\mathbf{T}$.

\begin{figure}[ht!]
\centering
\includegraphics[width=1\textwidth]{{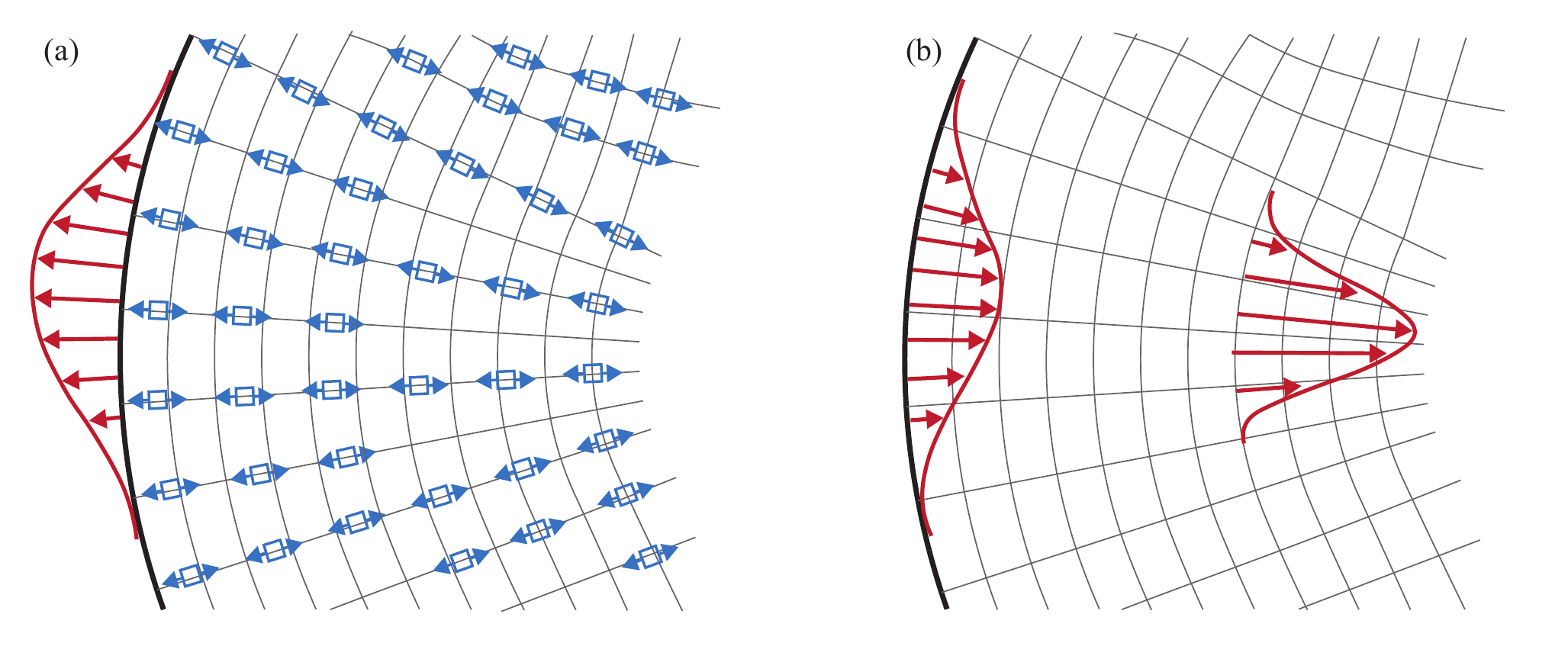}}
\caption{(a) An inverse problem showing part of a deformed body with a specified force distribution (red) on the boundary (thick black line), and strain distribution. The blue boxes show the measured principal directions of the strain. Because we assume the principal directions of $\mathbf{T}$ are the same as strain, characteristic lines can be drawn which specify the principal directions of $\mathbf{T}$ (gray lines). No shear stress may be transferred across these lines. (b) No shear means the force between lines is conserved. The stress just inside the boundary is equal and opposite. This stress is then amplified as the lines converge.}
\label{fig:line_convergence}
\end{figure}

We now consider how the solution procedure works in the two-dimensional example shown in Fig. \ref{fig:line_convergence}a. This specifies an initial problem with a force boundary condition and a map showing $\mathbf{\dot{E}}$. Using this $\mathbf{\dot{E}}$ field we can draw a grid of lines which are aligned the principal axis of $\mathbf{\dot{E}}$ at each point in the material. The lines will always be orthogonal at each point and define a curvilinear coordinate system. Since the principal directions of $\mathbf{T}$ are also aligned with the curvilinear coordinate system, no shear stress may be transferred across these lines. Since no shear is transferred, the force between two lines must be conserved and is specified at the boundary. Hence, when the lines converge the stress is amplified (Fig. \ref{fig:line_convergence}b). These lines are the characteristic lines of the governing hyperbolic partial differential equation.

We just assumed that there was no force propagation along the upward characteristic lines as propagation along curved characteristics is more complicated. Fig. \ref{fig:around_corner} shows the two-dimensional stress propagation along curved characteristics and the reduction in stress propagating in the radial direction. This interaction is captured by considering force balance in radial direction using polar coordinates $(r,\theta)$:

\begin{figure}[ht!]
\centering
\includegraphics[width=0.45\textwidth]{{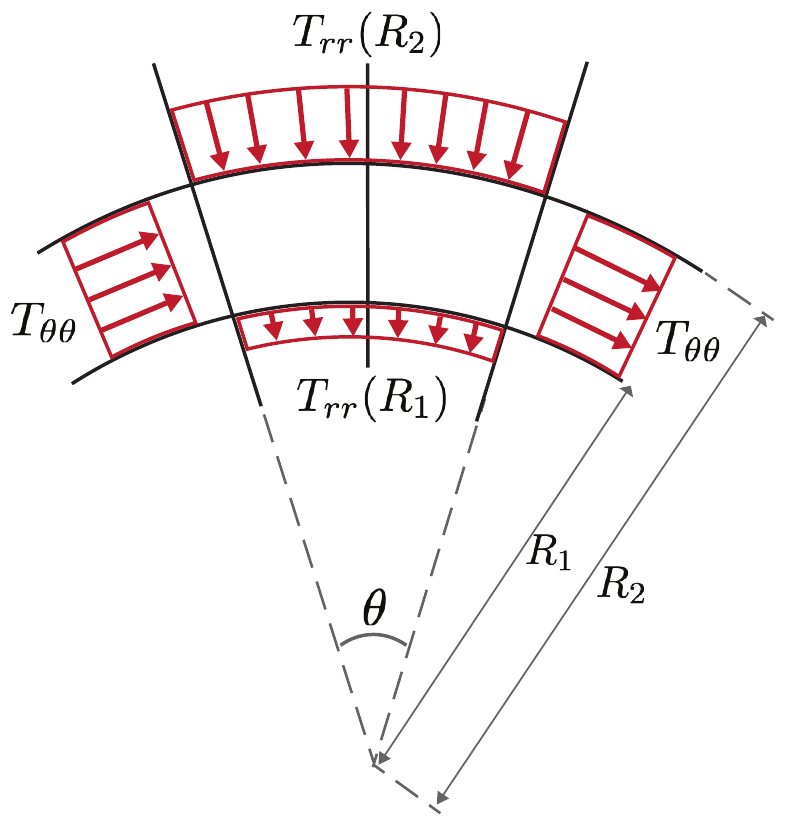}}
\caption{A constant stress $T_{\theta \theta}$ is propagating between two characteristic lines around a bend. This causes a change in the force propagating between characteristic lines in the radial direction. }
\label{fig:around_corner}
\end{figure}

\begin{equation}
    \frac{\partial}{\partial r}(rT_{rr}) + \frac{\partial T_{r\theta}}{\partial \theta} - T_{\theta \theta} = 0.
\end{equation}

Here the shear component $T_{r\theta}$ is zero since the coordinate system is chosen to be aligned with the characteristic lines. If $T_{\theta \theta}$ is set to be constant the force per unit depth propagating between two radial lines is $F_{rr}=T_{rr}r\theta$ and changes according to

\begin{equation}
    \frac{\partial F_{rr}}{\partial r} = \frac{T_{\theta \theta}}{\theta}.
\end{equation}

The general case is again more complex, with some curvature and divergence of characteristic lines in all three dimensions. However, this is quantitatively captured in Eqs. (\ref{eq:codir3}-\ref{eq:force3}). In this three dimensional case, the characteristic lines where shear cannot cross become surfaces \cite{Zachmanoglou1986}. This intuition regarding force propagation along characteristics gives insight into what boundary conditions are required in order to solve the governing equations. Specifically, the boundary stress only needs to be specified at one end of each characteristic line \cite{Zachmanoglou1986}.

\section{Range of applicability}
\label{sec:range_of_applicability}

Eqs. (\ref{eq:codir3}-\ref{eq:force3}) allow one to compute the full-field stress when (i) the boundary conditions are known, (ii) the deformation field is known, and (iii) the material is consistent with the alignment assumption. The authors anticipate several scenarios that meet these conditions where the approach will provide practical utility. First, in characterizing the strain softening of a material which is often impaired by deformation instabilities such as shear band formation or necking. Here, obtaining the full strain field and boundary conditions is often feasible using approaches such as digital image correlation and load cells \cite{Pan2009,Kang2005,Yan2015}. The computed stress field will allow one to characterize the material in the unstable region and could significantly aid in the characterization of granular materials \cite{Anand2000}, rubbers \cite{Feichter2007}, metals \cite{Zhao2016}, ceramics \cite{Rethore2007} and glasses \cite{Schroers2004}. Second, in characterizing material heterogeneities which may occur microscopically or microscopically, for example, graded materials and metal micro-structures \cite{Liu2018}. Digital image correlation can again be used in both these cases. An interesting example of heterogeneous behavior is turbulent fluid flow where the turbulent viscosity varies as a function of position and time \cite{Boussinesq1897} and could be determined using our approach based on data obtained with particle image velocimetry. Third, understanding material behavior during complex evolving deformations such as crack growth. Deformation data can be obtained from setups such as those described in \cite{Martinez2014}. It may even be possible to apply the approach to complex field observations such as landslides and geologic deformations, although determining the deformation field and boundary conditions are likely to be highly involved processes.

Despite its simplicity, the alignment assumption is valid for a range of materials with highly complex constitutive equations, allowing one to characterize these complexities using our approach. Examples include non-linear stress strain responses such as those that arise in hyperelastic deformation of elastomers \cite{Feichter2007} (e.g. see Fig. \ref{fig:one_dimension}); dependencies on variables such as density, temperature and strain rate; and dependencies on position in the material. Nevertheless, there are also simple cases where the approach breaks down such as idealized elasto-plastic deformation where one cannot distinguish $\mathbf{\dot{E}^e}$ and $\mathbf{\dot{E}^p}$. We do not attempt to provide an exhaustive list of when the approach does and does not apply as the number of materials and constitutive equations is vast. Instead, we select key examples of elasticity, associative flow rules and non-associate flow rules to provide general insights.
 
\subsection{Elasticity}
As discussed in \S \ref{sec:theory} the approach applies in the case of isotropic elasticity and is not restricted to the linear case. Complex non-linear hyperelastic behavior, such as that which occurs in rubbers or ceramics, is consistent with the alignment assumption. We also consider linear anisotropic elasticity which governs a number of materials ranging from crystals to wood. Here we have
\begin{equation}
    \mathbf{T} = \mathbb{C}\mathbf{E},
\end{equation}
where $\mathbb{C}$ is a fourth rank elasticity tensor. Although the symmetries of a material (e.g. that it is transversely isotropic) allow us to significantly reduce the number of coefficients in $\mathbb{C}$, they generally do not allow us to compute the principal directions of stress from the strain without further assumptions or data (unless it's isotropic). Nevertheless there are special cases. For example, when a transversely isotropic material is subject to plane strain or plane stress, this allows the components of stress in the plane to be computed. Another case is when the principal directions of strain are aligned with the symmetry axis of an orthotropic material. 

\subsection{Associative flow rules}
As an alternative to co-directionality, an associative flow rule may be used to specify the plastic flow direction $\mathbf{N^p}$. In this case, $\mathbf{N^p}$ is normal to the yield set and is given by

\begin{equation}
    \mathbf{N^p} = \frac{\mathbf{\dot{E}}}{|\mathbf{\dot{E}}|} = \frac{\partial f(\mathbf{T})}{\partial \mathbf{T}},
\end{equation}
where $f(\mathbf{T})$ is the yield function. For simplicity, we omit additional arguments such as the equivalent plastic tensile strain and assume a rigid-plastic material. The normality equation can be derived using using the maximum dissipation hypothesis \cite{Gurtin2010}. 

First we consider the classic case of the von Mises yield function
\begin{equation}
    f(\mathbf{T}_0) = \sigma - Y,
\end{equation}
where
\begin{equation}
    \sigma  = |\mathbf{T}_0|,
\end{equation}

and $Y$ is the yield stress \cite{VonMises1913}\footnote{For clarity of presentation we omit constant factors such as $\sqrt{3/2}$.}. The von Mises flow rule is classically derived using the co-directionality assumption, in which case our approach is valid. In the case where normality is used, this gives exactly the same result and our approach can still be applied.

Next we consider the Tresca yield criterion with the normality assumption. Unlike the von Mises criterion, this violates co-directionality \cite{Tresca1864,Coulomb1773}. Here the yield function is
\begin{equation}
    f(\mathbf{T}) = \textrm{max}(|T_1-T_2|,|T_1-T_3|,|T_2-T_3|)-Y,
\end{equation}
where $T_{i}$ are the principal components of $\mathbf{T}$.
If we express $f(\mathbf{T})$ in a reference frame aligned with its principal axis, the off diagonal terms of $\frac{\partial f(\mathbf{T})}{\partial \mathbf{T}}$ will be zero, hence, $\mathbf{\dot{E}}$ will have its principal directions aligned with $\mathbf{T}$. This can be extended to any $f(\mathbf{T})$ which is only a function of the principal components $T_{i}$. Hence, our approach applies to any associative flow rule where the yield function is isotropic (i.e. only a function of the stress invariants or principal components). Note that while the approach is valid for these associative flow rules with isotropic hardening, it breaks down in the case of kinematic hardening when the material is subject to a complex strain path. Here, only the principal directions of $\mathbf{T}- \mathbf{T_{back}}$ are known, where $\mathbf{T_{back}}$ is the back stress and remains unknown without further assumptions.

\subsection{Non-associative flow rule example - dry granular material}
Finally, we consider the example of isotropic dry granular material which gives rise to a non-associative flow rule that violates commonly used simplifying assumptions such as co-directionality, but for which our approach still applies. Here, the constitutive equations may be written as:

\begin{equation}
    \begin{split}
        \mathbf{D} =  \dot{\gamma}\left[\frac{\mathbf{P}+\beta \mathbf{N}}{|\mathbf{P}+\beta \mathbf{N}|}\right],\\
        \mathbf{P} = (1/2)\sin{(2\xi)}[\mathbf{\hat{e}}_1 \otimes \mathbf{\hat{e}}_1 - \mathbf{\hat{e}}_3 \otimes \mathbf{\hat{e}}_3],\\
        \mathbf{N} = \mathrm{sin}^2(\xi)\mathbf{\hat{e}}_1 \otimes \mathbf{\hat{e}}_1+\mathrm{cos}^2(\xi)\mathbf{\hat{e}}_3 \otimes \mathbf{\hat{e}}_3,
    \end{split}
\end{equation}
where $\dot{\gamma}$ is a strain rate scalar, $\mathbf{P}$ accounts for the material shearing, $\mathbf{N}$ accounts for the dilatancy of the material, $\xi$ is the slip plane angle, $\beta$ is the dilatancy parameter, $\mathbf{\hat{e}}_1$ is the direction of the maximum principal stress, and   $\mathbf{\hat{e}}_3$ is the direction of the minimum principal stress (see \cite{Anand2000} for more details and definitions associated with these constitutive equations). The key point here is that both $\mathbf{P}$ and $\mathbf{N}$ have principal directions aligned with $\mathbf{T}$ so our approach can be directly applied. Note that we assume the elastic strains and strain rates are small, which is frequently the case for granular materials.

This example also illustrates that our approach is valid for materials that violate the common assumption that plastic flow is independent of the pressure (e.g. \cite{VonMises1913}). This assumption is commonly applied to metals but is inaccurate in some cases \cite{Bridgman1953,Weir1998}.

\subsection{Cases that violate the alignment assumption}
\label{sec:allignment_violation}
While the emphasis has been on cases where the alignment assumption is valid, the only real requirement is that the principal directions of $\mathbf{T}$ are known. If there is some other orientation relationship that can be obtained from the deformation field this may be sufficient. For example consider an associative flow rule with an anisotropic yield function which depends on all components of $\mathbf{T}$ and does not adhere to the alignment assumption. If this yield function shape is known, i.e.

\begin{equation}
    \mathbf{N^p} = \frac{\partial f(\mathbf{T})}{\partial \mathbf{T}},
\end{equation}

but the hardening law and yield stress is unknown, one may still be able to apply our approach by inverting this equation, assuming it is one to one. This will give the principal directions of stress which can be substituted into Eqs. (\ref{eq:codir3}-\ref{eq:force3}).

There is also the important special case where the material does not generally follow the alignment assumption, but does for a specific deformation. For example, the case of elasto-plastic deformation where $\mathbf{E}$ and $\mathbf{\dot{E}}$ have the same principal directions and $\mathbf{E^e}$ cannot be neglected. This often occurs when the deformation is small because the minor changes in geometry are often not sufficient to significantly change the principal directions of $\mathbf{E}$. I.e. $\mathbf{E} = \mathbf{\dot{E}}t + \mathcal{O}(t^2) \approx \mathbf{\dot{E}}t$ when $t$ is small. Practically, this can checked for any specific finite deformation by simply observing if the principal directions of $\mathbf{E}$ change. 

To summarize, our approach can be applied whenever material symmetries or other assumptions allow us to deduce the principle directions of the stress from the deformation fields. While this can be applied directly to a range of isotropic materials, it does not apply in most anisotropic materials, materials with kinematic hardening subject to complex strain paths, or general elasto-plasticity. However, that it may be possible to extend this model with additional assumptions that overcome these issues.

\section{Analytical examples}
\label{sec:examples}
\subsection{Radially bent bar}
\label{sec:radially_bent_bar}
\begin{figure}[ht!]
\centering
\includegraphics[width=1\textwidth]{{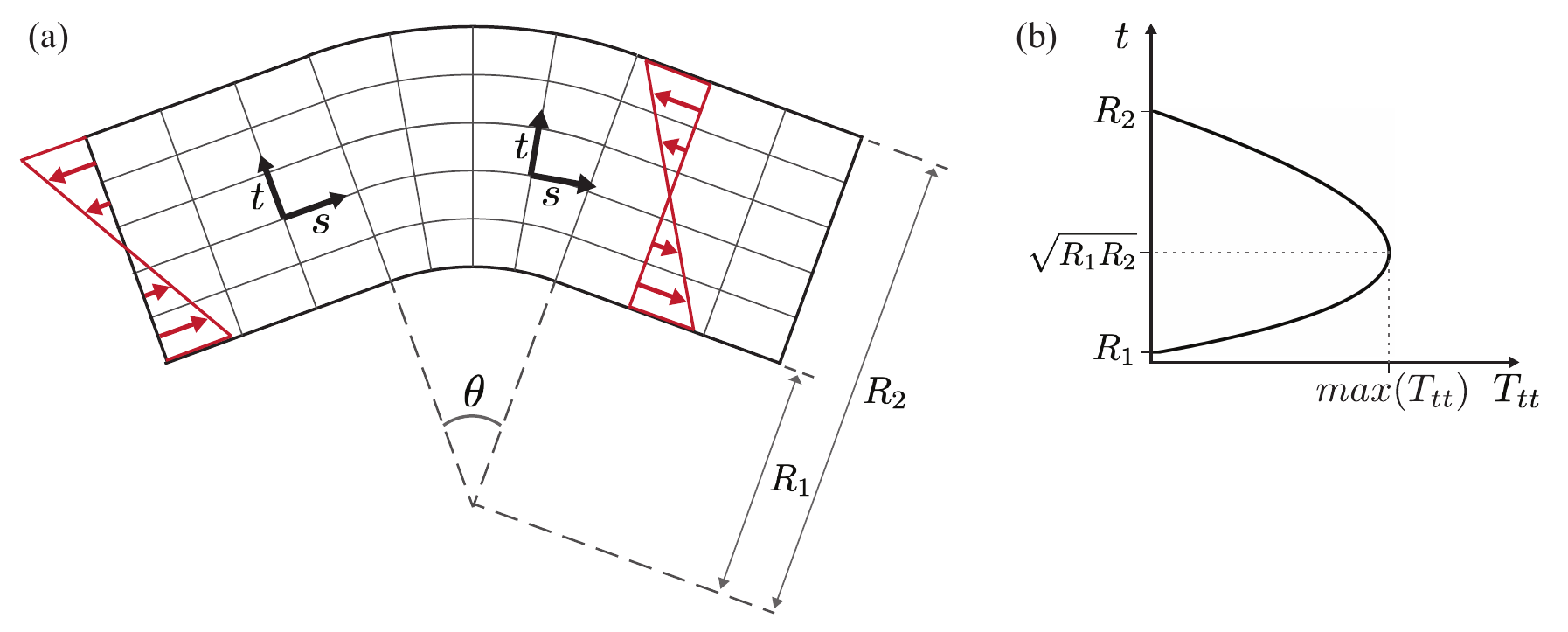}}
\caption{(a) An initially straight bar subject to a force distribution at each end and is plastically deformed in a radial manor in the center section. The characteristic lines (principal directions of $\mathbf{D}$) are specified by the curvilinear coordinate system $(s,t)$. (b) The analytical solution of $T_{tt}$ in the middle segment for the stress distribution shown in (a).}
\label{fig:bent_bar}
\end{figure}

Here we analytically examine the example of a bar plastically bending in the center, something that is commonly observed (Fig. \ref{fig:bent_bar}a) \cite{Huang2000,Lele2009}. An initially straight bar is subject to an applied traction normal to the boundary at each end. The traction at the other boundaries are set to zero. A strain distribution giving rise to the characteristic lines depicted in Fig. \ref{fig:bent_bar}a is assumed (this is consistent with those measured in \cite{Lele2009}). These characteristic lines define the curvilinear coordinate system $(s,t)$ used for our solution. We do not make the assumption of infinitesimal deformation, but require that the principal directions of $\mathbf{D}$ and $\mathbf{B}$ are aligned in accordance with Eq. (\ref{eq:codir_finite_strain}). However, as discussed earlier, the same analysis would apply to pure elastic deformation or pure plastic deformation. Here we specify that $T_{ss}=f(t)$ at the left boundary and $T_{tt}=0$ on the lower boundary.

Because the $t$ characteristic lines have no curvature, there will be no change in $T_{ss}$ in the $s$ direction (\S \ref{sec:stress_propagation}). Furthermore, $T_{st}=0$ since the principal directions are aligned with the curvilinear coordinate system. Hence, we have

\begin{equation}
    T_{ss}(s,t) = f(t), \qquad  T_{st}(s,t) = 0.
    \label{eq:barTsol1}
\end{equation}

$T_{tt}$ will also be equal to a constant in the straight segments of the bar, since there is no curvature in the $s$ characteristic lines. Due to the zero traction boundary condition on the lower side of the bar $T_{tt}(t=R_1)=0$, we have:

\begin{equation}
    T_{tt}(s,t)=0,
    \label{eq:barTsol2}
\end{equation}

for the straight segments. We now consider the bent segment using force balance in the $t$ direction (using standard polar coordinates):

\begin{equation}
    \frac{\partial}{\partial t}(tT_{tt})+T_{ss}=0.
\end{equation}

Hence, we have the following equation
\begin{equation}
    \int_{t={R_1}}^tT_{ss}dt= \int_{t={R_1}}^rf(t)dt= tT_{tt}\Big]_{R_1}^t.
    \label{eq:barTint1}
\end{equation}

Applying the zero traction boundary condition at the lower surface $T_{tt}(s,t=R_1)=0$ we have

\begin{equation}
    T_{tt}(s,t)=\frac{1}{t}\int_{t={R_1}}^tf(t)dt,
    \label{eq:barTint2}
\end{equation}
for the bent segment. This equation, combined with Eqs. (\ref{eq:barTsol1}-\ref{eq:barTsol2}), is the final form of the solution without specifying the specific traction distribution at the boundary $f(t)$. Note that no information from the left hand side or the top boundary has been used to solve the problem, and that these are actually computed as part of the solution. Furthermore, no information about the elasto-plastic constitutive equations is specified other than that the principal directions must be aligned. In order for our solution to be physically consistent with no traction on the top boundary, $f(t)$ is restricted to distributions which satisfy

\begin{equation}
    \int_{t={R_1}}^{R_2}f(t)dt=\int_{t={R_1}}^{R_2}T_{ss}dt = R_2T_{tt}(R_2)-R_1T_{tt}(R_1) =0,
    \label{eq:barTint3}
\end{equation}
because $T_{tt}(R_1)=0$ and $T_{tt}(R_2)=0$. Hence, the average axial stress must be zero. This is physically consistent with the observation that if this were nonzero, there would be a net vertical force acting on the bar and it would not be in equilibrium. 

We now consider a specific example of a force distribution:

\begin{equation}
    f(t) = T_{ss} = \sigma\left[\frac{(R_1+R_2)}{R_2-R_1}-\frac{2t}{R_2-R_1}\right].
\end{equation}

This changes linearly with $t$, takes on a value of $\sigma$ at $t=R_1$, takes on a value of $-\sigma$ at $t=R_2$, and satisfies Eq. (\ref{eq:barTint3}) (Fig. \ref{fig:bent_bar}a). Now we apply Eq. (\ref{eq:barTint2}) which gives

\begin{equation}
    T_{tt} = \sigma\left[\frac{t}{R_2-R_1}-\frac{R_2R_1}{t(R_2-R_1)}+\frac{R_1+R_2}{R_2-R_1}\right].
\end{equation}

The maximum value of $T_{tt}$ is:
\begin{equation}
    max(T_{tt}) = \sigma \left[\frac{2R_1R_2}{R_2-R_1}-\frac{R_1+R_2}{R_2-R_1}\right]
\end{equation}

which occurs at $t=\sqrt{R_1R_2}$. This relationship is depicted in Fig. \ref{fig:bent_bar}b.

\subsection{Fluid flow between two plates}
\label{sec:fluid_flow}
Here, show how this approach may also be applied to fluid flow. We have a two dimensional fluid flow between two parallel plates subject to a constant pressure gradient $-\frac{d P}{d x}$. Here we take an Eulerian approach and specify the velocity field $\mathbf{v}(\mathbf{x},t)$. We make the following assumptions: (i) the flow is steady, (ii) the flow is incompressible, (iii) the velocity is $x$ independent,  (iv) and the velocity is in the $x$ direction. This gives $\mathbf{v}(\mathbf{x},t) = v_x(y)$ where $x$ is in the direction of fluid flow and $y$ coordinate is perpendicular to the two plates.

This flow condition gives a velocity gradient tensor $\mathbf{D}$ with only off diagonal terms. Hence, using Eq. (\ref{eq:c}), $c$ will be equal to zero, and the principal directions of stress and strain will be oriented 45$^o$ from the horizontal (Eq.(\ref{eq:theta})). In this case, Eqs. (\ref{eq:forcex}-\ref{eq:codir2}) reduce to

\begin{equation}
    \frac{\partial T_{xx}}{\partial x} + \frac{\partial T_{xy}}{\partial y} = 0,
    \label{eq:fluid1}
\end{equation}

\begin{equation}
    \frac{\partial T_{xy}}{\partial x} + \frac{\partial T_{yy}}{\partial y} = 0,
    \label{eq:fluid2}
\end{equation}

\begin{equation}
    T_{xx}-T_{yy} = 0.
    \label{eq:fluid3}
\end{equation}

Solving these equations along with $T_{xx} = P$ gives: 
\begin{equation}
        T_{yy}=P,
\end{equation}
\begin{equation}
        \frac{\partial T_{xy}}{\partial y}= - \frac{\partial P}{\partial x}=\mathrm{constant}.
\end{equation}
This agrees exactly with the specific case of Poiseuille flow \cite{Kundu2012} which makes the assumption of constant viscosity. We did not make this assumption, and also did not need to make the additional assumption that $T_{xx}=T_{yy}=P$ which came from alignment of principal directions. Hence, this solution also applies to a visco-plastic flow.

\section{Conclusion}
\label{sec:conclusion}
Here we demonstrated the inverse problem can be formulated deterministically and solved exactly in a number of cases. This approach can be applied whenever the principal directions of stress are aligned with the principal directions of strain or strain rate. A proof of principle study is conducted showing the approach gives essentially exact results in the two dimensional case of spatially varying, plastic, constitutive relations. The sensitivity of the solution to this error in the input data was quantified. Further numerical validation may be beneficial for more complex three dimensional deformation problems, though we do not anticipate any particular issues as the equations are still relatively straightforward and linear. This alignment condition is readily realized in a range of material classes such as isotropic metal alloys, granular materials, polymers, and fluids, but not in textured or anisotropic materials, and materials with kinematic hardening subject to complex strain paths. However, it may be possible to extend the approach to these cases with additional assumptions that specify the principal directions of the stress based on the material and deformation path. We believe that the approach will have utility in experimental design and analysis in scenarios that involve complex geometries, plastic flow instabilities, or heterogeneous material properties. More importantly, it will enable straight-forward determination of constitutive relations: if the full field deformation is measured for multiple increments, one can directly compute the local stress-deformation path and fit a constitutive relationship to characterize the material.

\bibliographystyle{unsrt}  
\bibliography{references}  %%% Remove comment to use the external .bib file (using bibtex)..
\appendix
\setcounter{secnumdepth}{0}
\section{Appendix A: Numerical methods}
\subsection{Finite difference discretization}
In this appendix we provide further details on the finite difference discretization used in \S \ref{sec:proof_of_principle}. We take $i$ and $j$ to indicate the node number in the $x$ and $y$ directions respectively. $\psi_{i,j}$ corresponds to the $\psi$ value at $x=i\Delta x$ and y=$j\Delta y$ where $\Delta x$ and $\Delta y$ are the node spacings in the $x$ and $y$ directions. The finite difference grid is set to overlay the finite element grid for computational simplicity and to avoid unnecessary interpolation issues, hence there are 400 nodes in the $x$ direction and 100 nodes in the $y$ direction.

A second order central difference scheme is used to approximate the $x$ derivative:

\begin{equation}
    \frac{\partial^2 \psi}{\partial x^2} \approx [\psi_{i+1,j}-2\psi_{i,j} + \psi_{i-1,j}]/\Delta x^2.
    \label{eq:psi_disc_x}
\end{equation}

Due to the hyperbolic nature of the equation and the chosen boundary conditions, a backwards difference equation is chosen to approximate the second $y$ derivative: 

\begin{equation}
    \frac{\partial^2 \psi}{\partial y^2} \approx [\psi_{i,j}-2\psi_{i,j-1} + \psi_{i,j-2}]/\Delta y^2.
    \label{eq:psi_disc_y}
\end{equation}

This is accurate to first order in the grid spacing. The stencil giving the sum of Eqs. (\ref{eq:psi_disc_x}-\ref{eq:psi_disc_y}) is depicted in Fig. \ref{fig:discretization}b. Next we choose a backward difference approximation for the mixed partial derivative:

\begin{equation}
    \frac{\partial^2 \psi}{\partial x\partial y} \approx [\psi_{i,j}-\psi_{i,j-1} -\psi_{i-1,j}+ \psi_{i-1,j-1}]/[\Delta x \Delta y].
    \label{eq:psi_disc_xy}
\end{equation}

This is depicted in Fig. \ref{fig:discretization}c. The equation for each node $(i,j)$ is given by substituting Eqs. (\ref{eq:psi_disc_x}-\ref{eq:psi_disc_xy}) into Eq. (\ref{eq:numerical_problem}. The strain data at the specific $(x,y)$ location is used to compute the coefficient $c$ at each node using Eq. (\ref{eq:c}).

For the boundary nodes on the left ($x=0$), we simply apply  Eqs. (\ref{eq:psi_disc_x}-\ref{eq:psi_disc_xy}) and specify that $\psi_{i-1,j}=\sigma [y^2-l_yy]/2$ from Eq. (\ref{eq:numerical_problem}). Similarly for the right boundary nodes ($x=l_y$) we specify that $\psi_{i+1,j}=\sigma [y^2-l_yy]/2$. For the bottom boundary nodes we have two boundary conditions specifying the value and gradient of $\psi$. At ($y=0$) we specify that $\psi_{i,j-1}=0$, and $\psi_{i,j-2}=\sigma l_y\Delta y/2$. For the nodes at $y=\Delta y$, we specify that $\psi_{i,j-2}=0$.

This system of linear equations (one for each node) are combined into a single matrix equation of the form $Ax = b$ using python, where $x$ is a vector of $\psi$ values, $A$ is a matrix of coefficients multiplying those values and $b$ is a vector containing the forces at the boundaries. The system is then solved to give all $\psi$ values using the Scipy sparse linear algebra package \cite{Jones2001}. 

\subsection{Noise evaluation methodology}
The following methodology is implemented to investigate random, non-spatially correlated, error in the displacement field. We compute a noisy strain field from the displacement field and use this to replace the original strain field in Fig. (\ref{fig:methodology}b). We then compute the stress using the inverse problem approach and validate against the original stress field computed from the FEM solver. Specifically, we take the displacement field, directly exported from the commercial finite element software. As there are $400\times 100$ nodes, each node can be used to represent a subset in a DIC computation: i.e. the number of subsets/nodes will be consistent with a typical DIC computation. For each component of the displacement vector at each node, $\mathbf{u}$, random Gaussian noise with a specified standard deviation $\mathrm{std}(u'_i)$ is added. The gradient for each component of the displacement is computed using first order linear regression using all displacement points up a specified length $l_{\mathbf{u}\rightarrow\mathbf{E}}$ away in each direction. The strain is computed using the small strain approximation: $\mathbf{E} = (\mathrm{Grad}(\mathbf{u}) + \mathrm{Grad}(\mathbf{u})^T)/2$. This computation is done for every node, giving the strain at every node. The stress is then computed using the same approach as described as \S \ref{ssec:numerical_implementation} and the error is evaluated using the same approach as described in \S \ref{sec:validation}. This is done for a range of different values of $\mathrm{std}(u'_i)$, and a range of different $l_{\mathbf{u}\rightarrow\mathbf{E}}$. 

We also investigate the effect of random, non-spatially correlated, noise in the assumed boundary conditions. We specifically consider a stress distribution $T_{xx}(y)$ acting on the $x=0$ and $x=l_x$ boundaries, and do not consider a $T_{xy}$ component acting on the boundary. Adding a $T_{xy}$ component would be inconsistent with the specified direction of the principal stress at the boundary governed by $c$. Recall that without error we have $T_{xx}=T_0$. In addition, we add random Gaussian noise with a specified standard deviation $\mathrm{std}(T'_{xx})$ to each point. For simplicity, we do not add noise to the no-stress boundary condition on the $y=0$ and $y-l_y$ boundary and we use the same stress distribution on the $x=0$ and $x=l_x$ boundaries. We numerically integrate $T_{xx}(y)$ twice to obtain the boundary condition $\psi(y)$. Again, we use the same approach described in \S \ref{ssec:numerical_implementation}-\ref{sec:validation} to compute the stress and evaluate the error. 

A similar approach is taken to investigate the effect of systematic error in the boundary condition. We arbitrarily chose a linearly varying distribution $T_{xx}=\sigma + \delta T(1-2y/l_y)$, so that the stress is higher on for lower $y$ values, and lower for high $y$ values. $\delta T$ characterizes the deviation of the stress from its original value and can be related to the standard deviation of the stress using $\delta T = \sqrt{3}\mathrm{std}(T'_{xx})$. Again we numerically integrate the stress to obtain the $\psi(y)$ boundary condition and use the same approach to compute the stress and evaluate the error.

\end{document}